\documentclass[fleqn,usenatbib]{mnras}

\usepackage{newtxtext,newtxmath}

\usepackage[T1]{fontenc}

\DeclareRobustCommand{\VAN}[3]{#2}
\let\VANthebibliography\thebibliography
\def\thebibliography{\DeclareRobustCommand{\VAN}[3]{##3}\VANthebibliography}

\usepackage{graphicx}	
\usepackage{amsmath}	

\title[SZ effect in Cosmic Voids]{Cross-correlation of cosmic voids with thermal Sunyaev-Zel'dovich data}

\author[Li et al.]{
Gang Li,$^{1,2}$
Yin-Zhe Ma$^{3,4}$\thanks{Corresponding author: Y.-Z. Ma, \url{mayinzhe@sun.ac.za}}
Denis Tramonte,$^{5,1}$
Guo-Liang Li$^{1}$
\\
$^{1}$Purple Mountain Observatory, Chinese Academy of Sciences, Nanjing 210023, China\\
$^{2}$School of Astronomy and Space Science, University of Science and Technology of China, Hefei 230026, China\\
$^{3}$Department of Physics, Stellenbosch University, Matieland 7602, South Africa \\
$^{4}$National Institute for Theoretical and Computational Sciences (NITheCS), Stellenbosch University, Matieland 7602, South Africa \\
$^{5}$Department of Physics, Xi'an Jiaotong-Liverpool University, Suzhou 215123, China}

\date{Accepted XXX. Received YYY; in original form ZZZ}

\pubyear{2015}

\begin{document}
\label{firstpage}
\pagerange{\pageref{firstpage}--\pageref{lastpage}}
\maketitle

\begin{abstract}
We provide a measurement of the deficit in the Sunyaev-Zel'dovich Compton-$y$ signal towards cosmic voids, by stacking a catalogue of 97,090 voids constructed with BOSS-DR12 data, on the $y$ maps built on data from the Atacama Cosmology Telescope (ACT) DR4 and the \textit{Planck} satellite. 
We detect the void signal with a significance of $7.3\,\sigma$ with ACT and $9.7\,\sigma$ with \textit{Planck}, obtaining agreements in the associated void radial $y$ profiles extracted from both maps. The inner-void profile (for angular separations within the void angular radius) is reconstructed with significances of $4.7\sigma$ and $6.1\sigma$ with ACT and {\it Planck}, respectively; we model such profile using a simple model that assumes uniform gas (under)density and temperature, which enables us to place constraints on the product $(-\delta_{\rm v}T_{\rm e})$ of the void density contrast (negative) and the electron temperature. The best-fit values from the two data sets are $(-\delta_{\rm v}T_{\rm e})=(6.5\pm 2.3)\times 10^{5}\,\text{K}$ for ACT and $(8.6 \pm 2.1)\times 10^{5}\,\text{K}$ for {\it Planck} ($68\%$ C.L.), which are in good agreement under uncertainty. The data allow us to place lower limits on the expected void electron temperature at $2.7\times10^5\,\text{K}$ with ACT and $5.1\times10^5\,\text{K}$ with \textit{Planck} ($95\%$ C.L.); these results can transform into upper limits for the ratio between the void electron density and the cosmic mean as $n^{\rm v}_{\rm e}/\bar{n}_{\rm e}\leqslant 0.73$ and $0.49$ ($95\%$ C.L.), respectively. Our findings prove the feasibility of using tSZ observations to constrain the gas properties inside cosmic voids, and confirm that voids are under-pressured regions compared to their surroundings.
\end{abstract}




\section{Introduction}

\label{sect:intro}

Cosmic voids are the largest under-dense regions filling most of the volume in the Universe, and 
contain abundant cosmological information for probing dark energy, cosmic structure growth, and 
galaxy formation~\citep{Lavaux2012,Pisani2015,Cai2015,Beygu2017,Falck2018,Pustilnik2019,Hamaus2020,Aubert2022,Ceccarelli2022}. Nonetheless, in general, void properties have not been investigated sufficiently, as 
their low-density environment makes it challenging to find observational tracers. In recent years, 
there have been several studies aimed at cross-correlating voids with other large-scale structure 
(LSS) tracers. For example, voids have been detected by using different LSS gravitational lensing 
data~\citep{Melchior2014,Clampitt2015,Sanchez2017,Fang2019}, and CMB lensing maps~\citep{Cai2017,Raghunathan2020,Vielzeuf2021}. However, because the lensing signal is only sensitive to the total matter density, 
cross-correlation with lensing data cannot dissect the baryonic gas from the underlying dark matter 
distribution.  

Characterising gas inside voids is indeed a crucial problem in cosmology, as voids can possibly host 
the ``missing baryons'' in the galactic and super-galactic ecosystem~\citep{Haider2016,Martizzi2019}, 
which are particularly challenging to detect\footnote{For a general description of the missing baryon 
problem, please refer to~\citet{Fukugita2004, Bregman2007} and~\citet{Shull2012}.}. Recent observations 
from O\uppercase\expandafter{\romannumeral7} absorption systems~\citep{Nicastro2018} and fast radio 
burst~\citep{Macquart2020} showed that most of the undetected baryons are likely in form of a diffuse 
and ionised plasma, such as the cool phase of the inter-galactic medium (IGM, $T<10^{5}\,{\rm K}$) 
and the warm-hot intergalactic medium (WHIM, $10^{5}<T<10^{7}\,{\rm K}$). However, because of the 
low-density ($n_{\rm H}<10^{-4}(1+z)\,{\rm cm}^{-3}$) of this diffuse gas, it is difficult to trace 
out the baryons location and pin down their detailed physical properties. Cosmological hydrodynamical 
simulations suggest that a significant portion of baryons is indeed contained within cosmic voids, in 
a multi-phase gas state~\citep{Haider2016,Martizzi2019}. It is therefore necessary to carry out a 
detailed observational study of gas inside cosmic voids.

In this work, we use the thermal Sunyaev-Zel’dovich (tSZ) effect~\citep{Sunyaev1970} to probe the 
warm-hot gas that is possibly contained inside cosmic voids. The tSZ effect arises from the inverse 
Compton scattering of CMB photons by the warm-hot ionized electrons in the LSS. The amplitude of tSZ 
effect is captured by the Compton-$y$ parameter, which is the integrated electron pressure along the 
line-of-sight (LoS): 
\begin{eqnarray}
y=\frac{\sigma_{\rm T}}{m_{\rm e}c^{2}}\int P_{\rm e} \, {\rm d}l,
\end{eqnarray}
where $P_{\rm e}\equiv k_{\rm B}n_{\rm e}T_{\rm e}$ is the gas pressure ($n_{\rm e}$, $T_{\rm e}$ 
being the electron density and temperature and $k_{\rm B}$ being the Boltzmann constant), 
$\sigma_{\rm T}$ is the Thomson cross section, $m_{\rm e}$ is the electron mass, and $c$ is the 
speed of light. In recent years, several works have been carried out to extract the WHIM's tSZ 
effect from cosmic filaments~\citep{Tanimura2019,deGraaff2019}, intercluster gas~\citep{Planck2013, Bonjean2018}, and correlated gas in and between dark matter 
halos~\citep{Ma2015,Hojjati2015,Hojjati2017,Makiya2018,Ma2021,Ibitoye2022,Pandey2022,Tramonte2023}. More 
importantly,~\citet{Alonso2018} reported the first detection of tSZ signal towards cosmic voids at $3.4\sigma$ 
confidence level (C.L.), by stacking SDSS CMASS voids\footnote{\citet{Alonso2018} void catalogue 
is constructed from the SDSS-DR12 CMASS galaxy sample, and can be downloaded 
at~\url{http://lss.phy. vanderbilt.edu/voids/}.} on the {\it Planck} Compton-$y$ map. The results 
indicate that voids are under-pressured relative to the cosmic mean. Such a detection corroborates 
the relevant abundance of warm-hot gas in voids and their surroundings. In order to further 
investigate the gas properties inside voids, we need to obtain more precise measurements of its 
density and temperature, and compare them with the theoretical modelling. This is the aim of this work. 

In the following, we will investigate the gas inside voids by using Compton-$y$ maps from both the {\it Planck} 2015 data release and the Atacama Cosmology Telescope (ACT) Data Release 4 (DR4), where the latter has a higher angular resolution than the former. In addition, we will use a void catalogue constructed with a novel
parameter-free cosmological void finder~\citep{Zhao2016}, which renders in total $97,090$ voids over the ACT 
footprint; this sample is $\sim125$ times the size of the CMASS-based void sample used in~\citet{Alonso2018}. 
Besides stacking these
voids on the $y$ maps, we will build a gas density and temperature model to interpret the measurement 
of the reconstructed mean void Compton-$y$ profile.

The rest of the paper is organized as follows. Sec.~\ref{sect:data} describes the Compton-$y$ maps 
and void data we use. Sec.~\ref{sect:analysis} presents the stacking method, results and the inclusion 
of possible CIB contaminations; we then build a theoretical model to quantify the void gas characteristics 
in Sec.~\ref{sect:gas}. We summarise our findings and discuss their physical implications in 
Sec.~\ref{sect:conclusion}. Throughout this work, we adopt a spatially-flat $\Lambda$CDM cosmology with 
the relevant parameters fixed to the values $h=0.68$ and 
$\Omega_{\rm b}h^{2}=0.0224$~\citep{Planck-2018parameters}.


\section{Data}
\label{sect:data}

\begin{figure*}
\centerline{
\includegraphics[width=9cm, height=3.0cm]{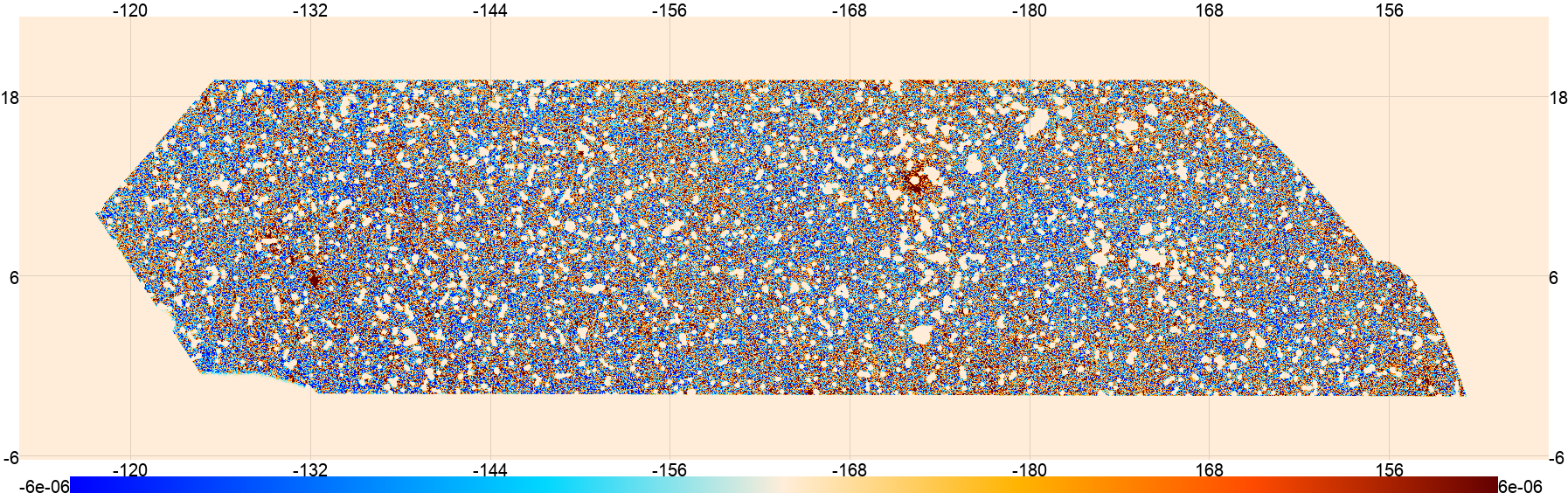}
\includegraphics[width=9cm, height=3.0cm]{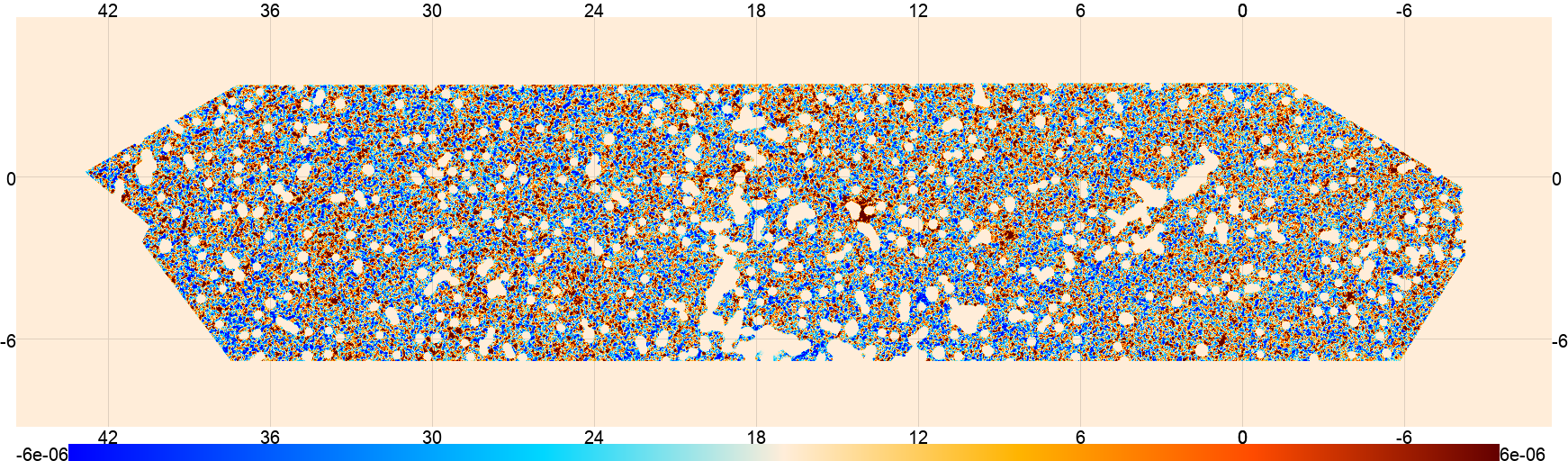}
}
\vspace{0.5cm}
\centerline{
\includegraphics[width=9cm, height=3.0cm]{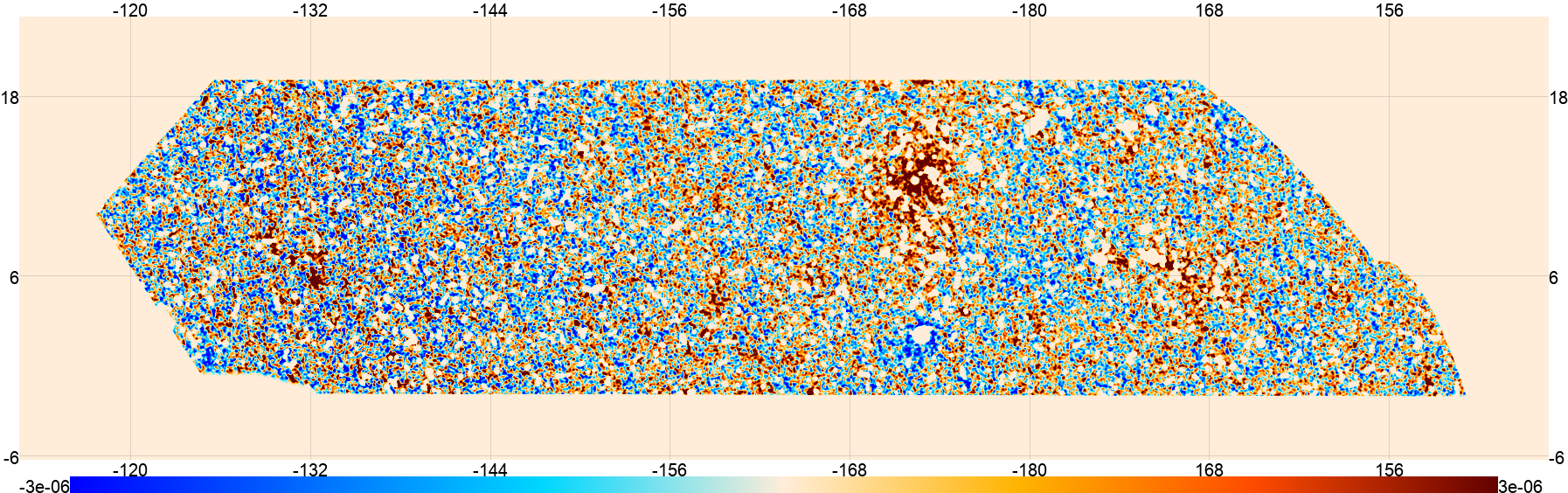}
\includegraphics[width=9cm, height=3.0cm]{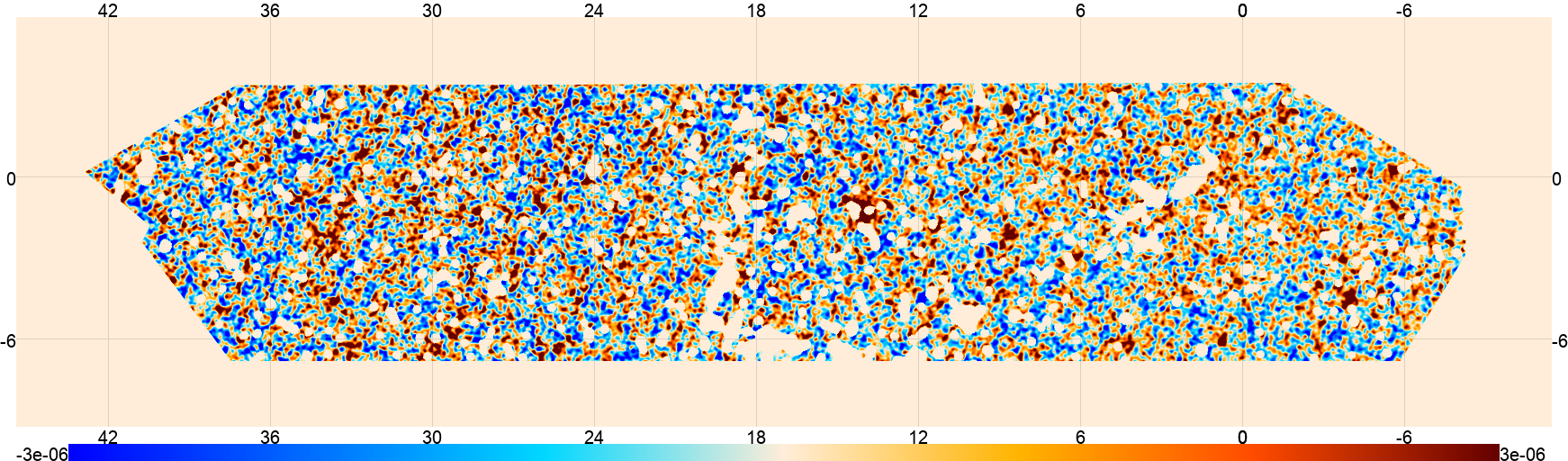}
}
\caption{Combined Compton-$y$ maps with the projected {\it Planck} Galactic Plane and point sources masks. The {\it left} column is the ACT BN-region, and the {\it right} is ACT D56 region. The {\it upper} row shows the actual ACT data and the {\it lower} row shows the {\it Planck} data on ACT region. The vertical axis (declination) and
horizontal axis (right ascension) are labeled in degrees.}
\label{fig: act-ymaps}
\end{figure*}

\subsection{Compton-$y$ maps} 

We use the Compton-$y$ maps obtained from ACT DR4 data that is presented in~\citet{Madhavacheril2020}. 
The ACT Compton-$y$ maps are the first wide-area ($\sim 2,100\,{\rm deg}^{2}$), arc-minute resolution 
component separated maps of the tSZ effect. They are constructed by implementing a component 
separation method based on the Internal Linear Combination (ILC) approach~\citep{Remazeilles2011}, 
combining \textit{Planck} individual frequency maps from 30 to 545 GHz and ACT data at 98 and 150 GHz. 
The ACT Compton-$y$ maps have an unprecedented high resolution of ${\rm FWHM}=1.6$ arc-minutes, and 
cover two separated regions, namely the D56 region ($\sim 456\,\text{deg}^2$), and the larger BN 
region ($\sim 1,633\,\text{deg}^2$).

A number of potential foreground contaminants such as the Galactic plane or radio sources may 
affect the cross-correlation analysis between the tSZ maps and other LSS tracers~\citep{Madhavacheril2020}. 
While the ILC pipeline generally does not fully eliminate the foreground residuals, there is also 
no specific mask released by the ACT collaboration that is tailored to tSZ analyses; hence, we adopt 
the \textit{Planck} $40\%$ Galactic plane mask combined with the \textit{Planck} point-source 
mask~\citep{Planck2016SZ-sources}, and project it on the ACT footprint (upper panels in Fig.~\ref{fig: act-ymaps}). 

Apart from the ACT $y$-maps, another available full-sky Compton-$y$ map was provided as part of 
\textit{Planck} 2015 data release~\citep{Planck-ymap2016}. The \textit{Planck} $y$-map is pixelized on 
a 2D spherical surface in HEALPix format with resolution $N_{\rm side}=2048$~\citep{Gorski2005}. 
There are two versions of the {\it Planck} $y$-map, based on different reconstruction methods: the Modified 
Internal Linear Combination Algorithm~\citep[MILCA, ][]{Hurier2013} and the Needlet Internal Linear 
Combination~\citep[NILC,][]{Remazeilles2011}; both are derived from multi-band combinations of 
\textit{Planck} intensity maps from 30 to 857 GHz. The NILC $y$-map has higher noise on large scales 
compared to the MILCA $y$-map, although it has been found that the former basically gives results 
consistent with the latter even for large-scale studies~\citep{Vikram2017,Alonso2018,Tanimura2019}. 
Nonetheless, since voids usually subtend large angular scales, to avoid potential noise contaminations, 
we will only employ the MILCA $y$-map for our void stacking analysis.

The {\it Planck} $y$-map has an angular resolution of ${\rm FWHM}=10$ arc-minutes, lower than the ACT $y$-map.
In order to allow a better comparison with the ACT $y$-map results, we first project the \textit{Planck} 
$y$-map on the ACT footprint regions using the {\tt pixell} package\footnote{\url{https://pixell.readthedocs.io/en/latest/}}, which also ensures the same stacking pipeline can be used for both versions of the 
Compton-$y$ map. The same mask is also used for both \textit{Planck} and ACT $y$-maps.

The final masked ACT and \textit{Planck} $y$ maps that we are going to employ in this study are shown in the top and bottom panels of Fig.~\ref{fig: act-ymaps}\footnote{ACT footprint with the projected masks is also shown in Fig.~1 of~\citet{Tramonte2023}.}.

\subsection{Void catalogue} 

We use the void catalogue constructed by~\citet{Zhao2016}, who employed a novel parameter-free 
cosmological void finder (DIVE, Delaunay TrIangulation Void findEr) based on a Delaunay Triangulation 
(DT) algorithm. The DIVE algorithm can efficiently compute the empty spheres constrained by a discrete 
set of tracers; hereafter we shall label the resulting voids as {\it DT voids}. The DT void catalogue is 
constructed from the BOSS-DR12 galaxy data, which include both CMASS (a subsample of 
BOSS galaxies with redshift $z>0.45$) and LOWZ (a subsample of BOSS galaxies with redshift $z<0.45$) 
galaxies~\citep{Alam2017,Tanidis2021}. The DT voids span the redshift range $0.1 < z < 0.8$, and the (comoving) radius range 
$1.0h^{-1}\text{Mpc} <R_{\rm v}< 99.3h^{-1}\text{Mpc}$; the full DT void catalogue contains in total 
7,886,816 voids, of which 5,713,918 in the northern sky and 2,172,898 in the southern sky. 

In our analysis we will only consider voids with redshift in the range $0.2<z<0.4$ and radius in the comoving
range $15.0h^{-1}\text{Mpc}<R_{\rm v}<25.0h^{-1}\text{Mpc}$; such a selection can indeed ensure a sufficient 
statistics for the stacking analysis, while avoiding uncertainties in the reconstructed void properties 
deriving from the boundaries of the catalogue. We then select the voids that are completely overlapping with the ACT $y$-map footprint, ensuring that the 
separation between their centres and the map boundaries is not smaller than two times their projected 
angular radius. After these steps, there are 81,053 and 16,037 voids left in BN and D56 regions, respectively. 
The mean redshift and effective (comoving) radius of the 97,090 selected voids are $z=0.32$ and 
$18\,h^{-1}{\rm Mpc}$, which corresponds to the effective physical radius $R_{\rm eff}=13.6\,h^{-1}{\rm Mpc}$.

\citet{Zhao2016} also generated $1.4\times 10^{8}$ mock voids ($4\times 10^{7}$ voids in 
the southern sky, $10^{8}$ voids in the northern sky), which we shall employ for the statistical 
characterisation of our results. Again, we first select the mock voids that 
satisfy the redshift and radius conditions detailed above, which leaves 9,928,676 mock voids in total. 
The subsequent positional query leaves 1,522,072 mock voids over the BN region and 317,829 mock voids 
over the D56 region. Finally, we construct 1000 different mock catalogues for each region, by randomly 
choosing each time 81,053 and 16,037 mock voids out of these BN and D56 mock samples, 
respectively. The mock catalogues are used to deduct the background signal as well as to compute 
the covariance matrix and the significance of our measurements, as detailed in Sec.~\ref{sect:analysis}.


\section{Measurements}
\label{sect:analysis}

\begin{figure*}
\centering{
\includegraphics[width=8cm]{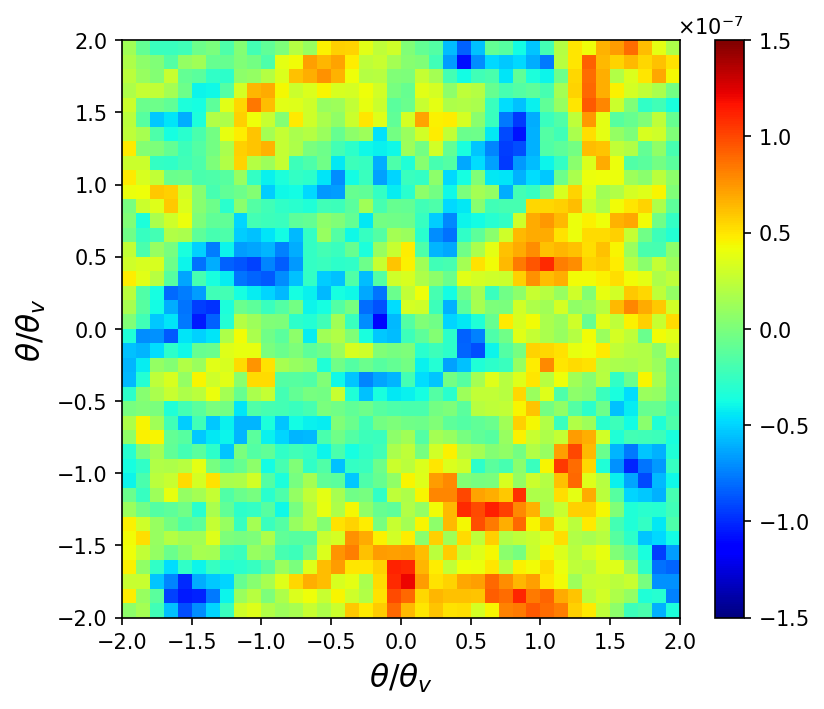}
\includegraphics[width=8cm]{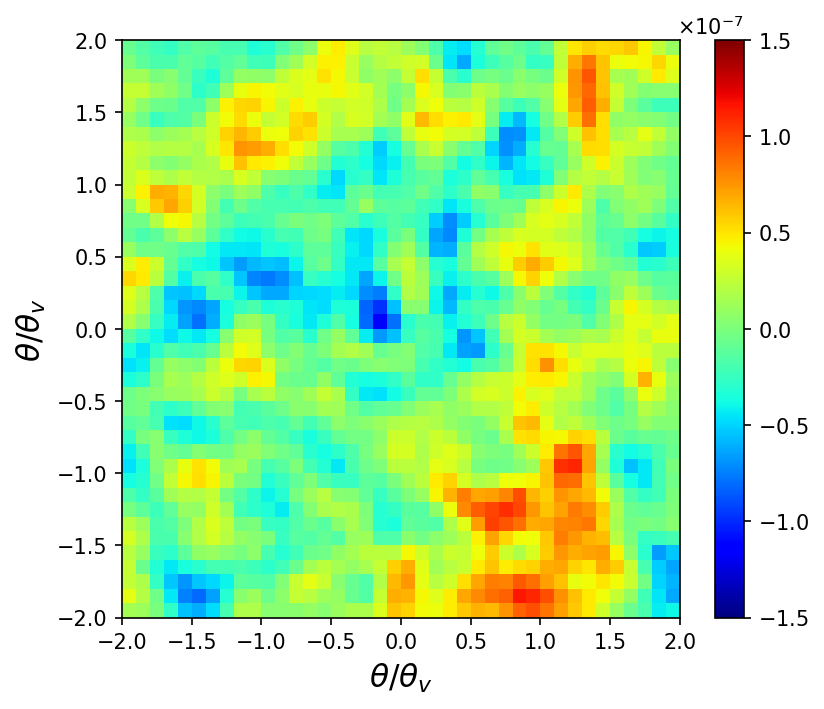} 
\includegraphics[width=8cm]{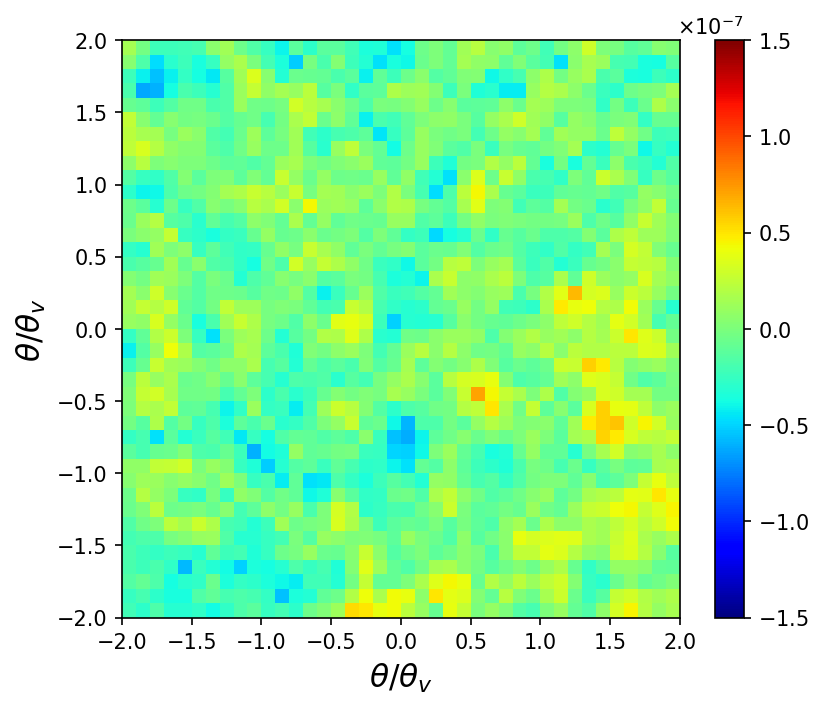}
\includegraphics[width=8cm]{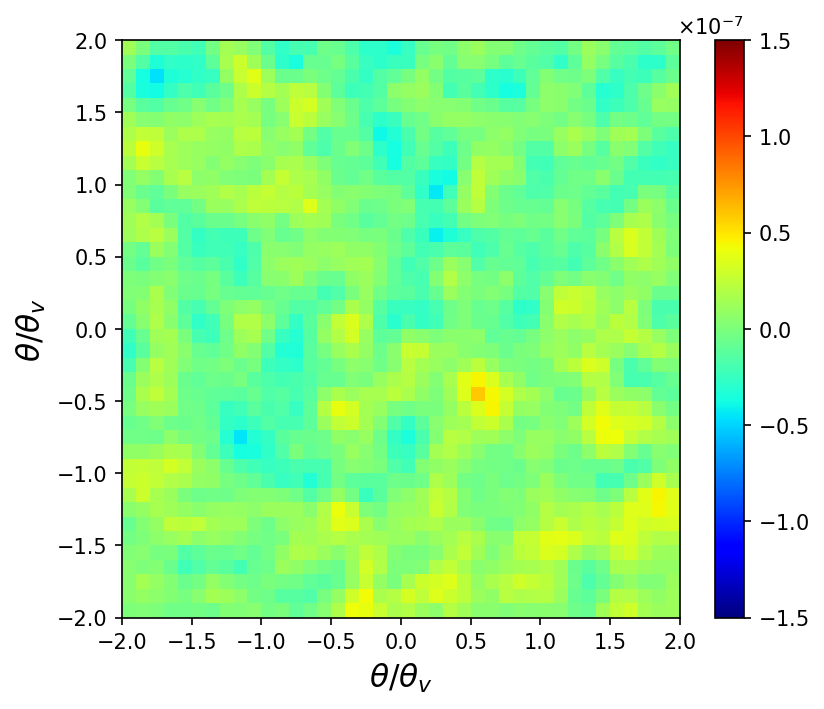} 
}
\caption{{\it Top:} the 2D signal maps obtained by stacking the DT void sample on ACT ({\it left}) and 
	{\it Planck} ({\it right}) Compton-$y$ maps. {\it Bottom:} same as in the top row, but showing 
	this time the results obtained using a random (mock) void catalogue. }
\label{fig: stack-results}
\end{figure*}

\begin{figure}
\centering{
\includegraphics[width=8.5cm, angle=0]{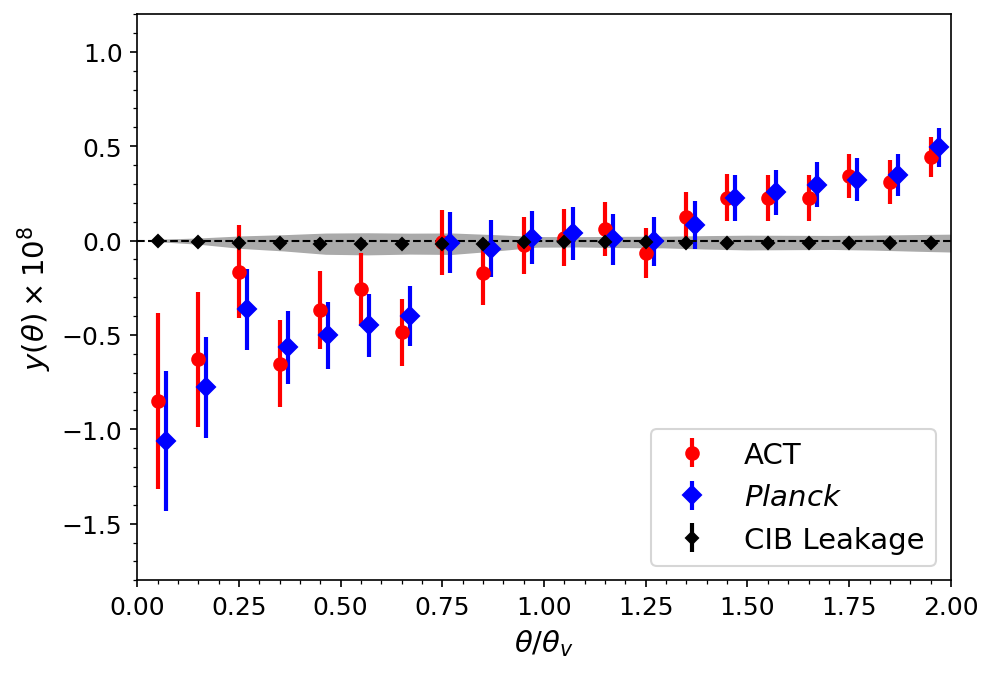}
}
	\caption{Reconstructed mean radial Compton-$y$ profile for the DT voids measured on ACT 
	(red circles) and {\it Planck} (blue squares) stacked maps (top row of 
	Fig.~\ref{fig: stack-results}). The black diamonds show the best-fit CIB leakage in the 
	stacked void profile, by scaling the stacked void signal on the Planck $545\,{\rm GHz}$ 
	temperature map with the $\alpha_{\rm CIB}$ parameter. The grey band around the zero line 
	shows the range of CIB leakage by allowing $\pm 1\sigma$ uncertainty on $\alpha_{\rm CIB}$. The estimation of $\alpha_{\rm CIB}$ is described in Sec.~\ref{sect:systematic}. The total significance of the detection is $7.3\sigma$ for ACT and $9.7\sigma$ for {\it Planck}, while the inner profile (considering only the points for which $\theta/\theta_{\rm v}\leqslant 1$) is detected at $4.7\sigma$ with ACT and $6.1\sigma$ with {\it Planck}.}
\label{fig:profiles}
\end{figure}

\begin{figure*}
\centering{
\includegraphics[width=6.5cm]{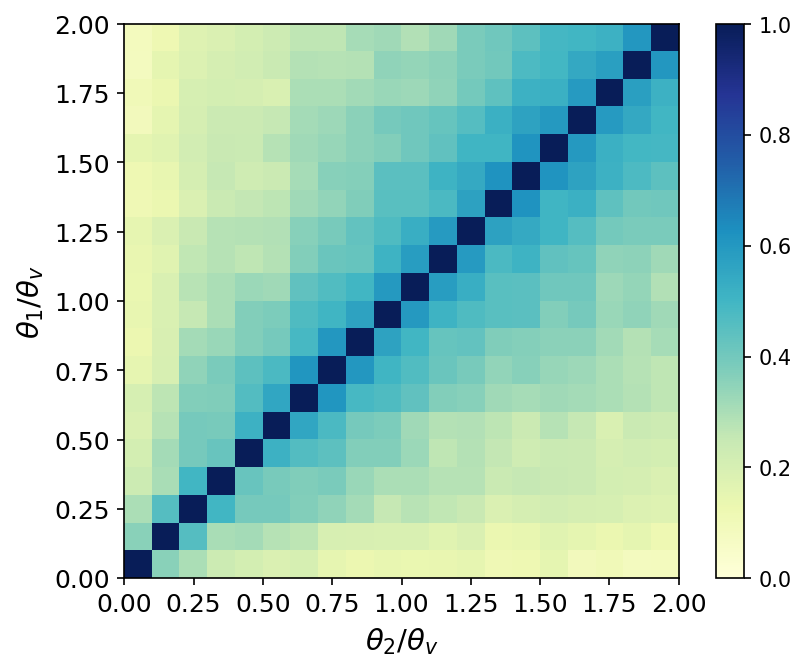}
\includegraphics[width=6.5cm]{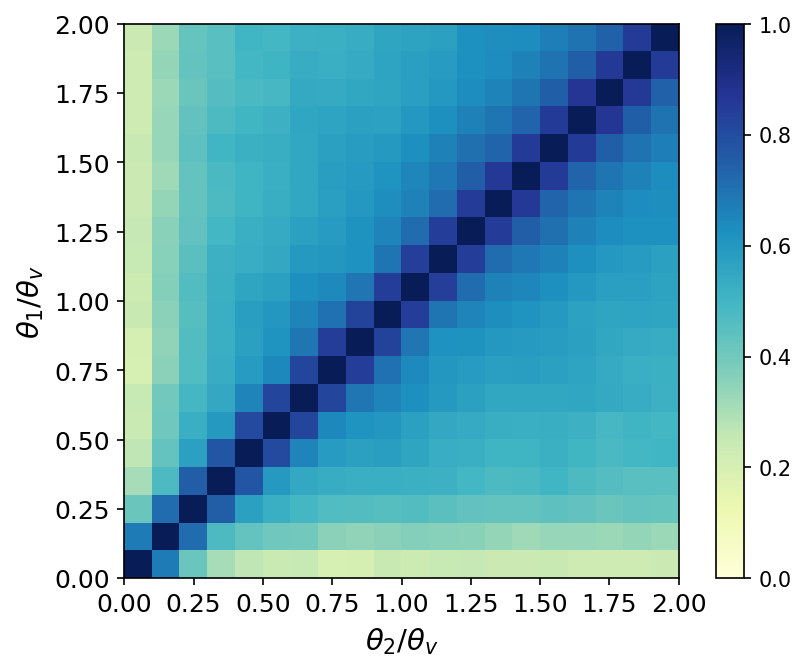}
}
\caption{The correlation matrices (Eq.~(\ref{eq:def_corr})) for the measured radial tSZ 
	profile around voids, for both the case of ACT (\textit{left}) and \textit{Planck} 
	(\textit{right}) $y$ maps. These correlation matrices are estimated from the stack 
	results of 1,000 mock void catalogues.} 
\label{fig: Corre-Matrix}
\end{figure*}

\begin{figure}
\centering{
\includegraphics[width=7.2cm]{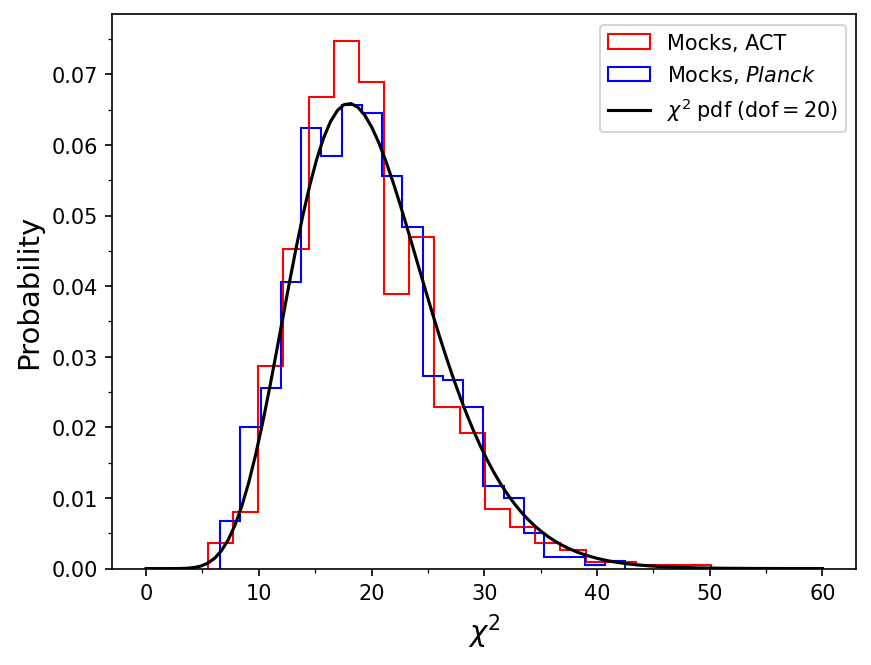}
}
\caption{The probability distribution of $\chi^{2}$ values for the 1,000 mock void catalogues, 
	plotted as a histogram for both ACT and \textit{Planck} results. 
	Because the mock voids and the $y$ maps are uncorrelated, this distribution is very 
	close to a ``chi-squared'' distribution with 20 degrees of freedom, which is also overplotted 
	as the solid black line. } 
\label{fig: chi2-mocks-v2}
\end{figure}

\begin{figure}
   \centering
   \includegraphics[width=7.2cm, height=5.3cm]{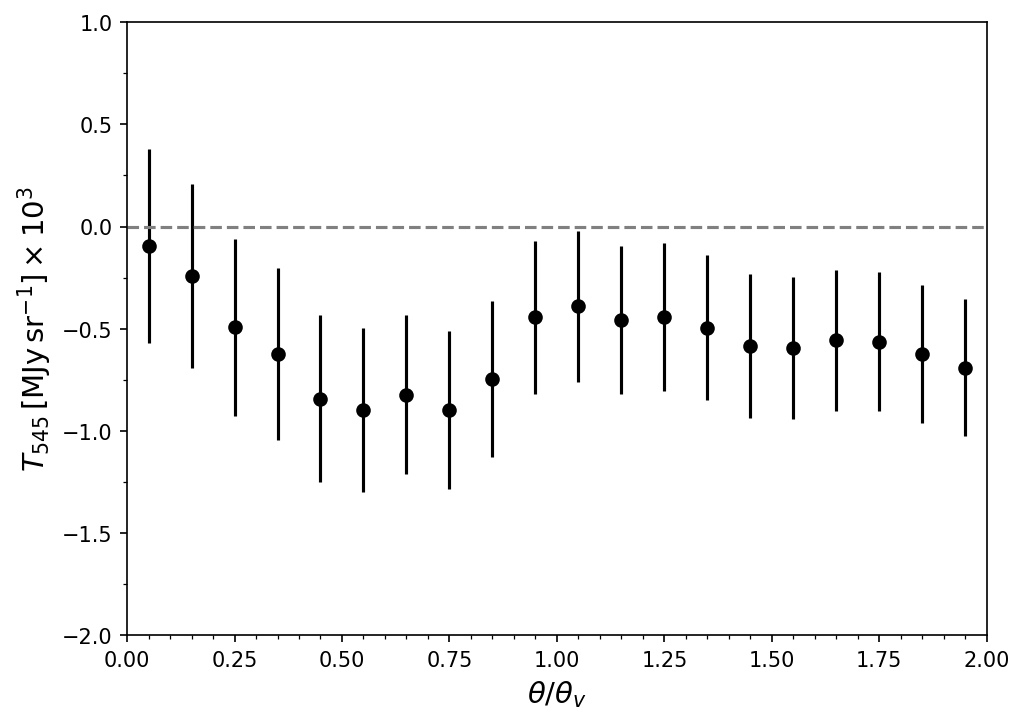}
   \caption{The stacked void profile on {\it Planck} 545 GHz temperature map, which can be regarded 
	as a proxy for CIB contamination to our $y$ measurement, when scaled by the 
	$\alpha_{\rm CIB}$ factor (see also Fig.~\ref{fig:profiles}).}
   \label{fig: T545-profile}
\end{figure}

\subsection{Stacking analysis} 

To extract the tSZ signal around voids, we stack the Compton-$y$ map at the position of the 
selected voids. The $y$-maps are pixelized on a 2D flat plane and we use the {\tt pixell} {\tt Python}
package\footnote{\url{https://pixell.readthedocs.io/en/latest/}} to extract all pixels within a square 
area centred on each void and with side equal to four times the void angular radius, $4\theta_{\rm v}$, 
where $\theta_{\rm v}\equiv R_{\rm eff}/D_{\rm A}$, with $D_{\rm A}$ the angular diameter distance to the 
void. We then scale the angular separation of each pixel from the void central 
pixel by the void effective radius ($\theta/\theta_{\rm v}$); this operations ensures that, prior to the 
stack, all voids are brought to a uniform size. The scaled pixels are then binned into the same grid 
coordinate system of $-2<\Delta \text{RA}/\theta_{v}<+2$ and $-2<\Delta \text{Dec}/\theta_{v}<+2$, 
divided into $40\times40$ bins. For the generic $i$-th void, this procedure is repeated for both the 
Compton-$y$ map, yielding a void signal map $y_i$, and for the mask, yielding a void weight map
$W_i$ (the latter only has possible values 0 and 1). For the full void sample, the resulting stacked 
signal map is then obtained as the weighted average:
\begin{eqnarray}
	y_{\rm stack} = \frac{1}{W_{\rm stack}}\sum_{i=1}^{N_{\rm v}} y_i\,W_i,
\end{eqnarray}
where $N_{\rm v}$ is the number of stacked voids, and 
\begin{eqnarray}
W_{\rm stack} = \sum_{i=1}^{N_{\rm v}} W_i 
\end{eqnarray}
is the total weight for this stack.

We obtain this way two different stack maps, one for each of the two ACT patches. 
The final, combined map can be obtained with a further weighted average as:
\begin{eqnarray}
    y_{\rm tot} = \left[\sum_{i=1}^{N_{\rm BN}} y^{\rm BN}_{i}\,W_{i}^{\rm BN}+\sum_{i=1}^{N_{\rm D56}} y_{i}^{\rm D56}\,W_{i}^{\rm D56}\right]\Big/
    \left[W_{\rm stack}^{\rm BN} +W_{\rm stack}^{\rm D56}\right].
    \label{eq:y_tot_v}
\end{eqnarray}


The procedure described above is also repeated for each of the mock catalogues, thus yielding 
$1,000$ mock $y$-stack maps; we label $y^{(k)}$ the stacked signal map for the $k$-th mock catalogue. 
These mock stacks are representative of the background contribution to the signal, which includes 
additional effects such as incomplete sky coverage, large-scale monopole contribution and residual 
systematics. Indeed, because we are stacking both the real void catalogue and the mock 
catalogues on the same map, any of such systematic effects affecting the latter will be captured by 
both the catalogue stack and the mock stacks. Our final reconstruction of the void signal is then obtained 
by subtracting from the real void stack, the average of the mock stack, which we can consider as 
a ``background'' term calculated as:
\begin{eqnarray}
	\label{eq:background_subtract}
	\bar{y} = y_{\rm tot} - \frac{1}{N_{\rm m}}\sum_{k=1}^{N_{\rm m}}y^{(k)}, 
\end{eqnarray}
where $N_{\rm m}=1000$ is the number of mock samples. We suppress the overhead bar in $\bar{y}$ for abbreviation in the following. A similar approach was also adopted in~\citet{Alonso2018}.

In Figure~\ref{fig: stack-results} we show the 2-dimensional stack maps, while the corresponding 
1-dimensional radial profiles are plotted in Fig.~\ref{fig:profiles}. Because the 2-D map are pixelised in $40\times 40$ grids, we use $20$ racial bins to compute the average $y$ value within each annulus of the 2-D map.
A signal amplitude 
deficit can be seen in the region $\theta/\theta_{v}<1$ both in the stacked 2D maps and also 
(more clearly) in the 1D profile. For comparison, Figure~\ref{fig: stack-results} also shows the 
stack maps obtained with one of the mock void catalogues: we can see that there is no central deficit 
when stacking the random catalogues (stacks for other mocks are qualitatively similar to the one 
shown here), which supports the interpretation of the central amplitude decrease,  when stacking 
the real catalogue, as void signal. One can also appreciate similar patterns in both the 2D  
ACT and \textit{Planck} stack maps using real voids, as well as similar trends in the associated
1D profiles. Figure~\ref{fig:profiles} shows the larger errors near the center of the stacked voids than outer radii. This comes from the fact that there are larger fluctuations in the center of the voids (the negative peak amplitude) compared to the outskirts where an average of zero level is more consistent from void to void. As a result, the smaller fluctuations in the outskirts and higher fluctuations in the center produce variances of the errors in different radii.

Finally, we notice that the void profile amplitude approaches zero at the unit scaled void 
radius ($\theta/\theta_{\rm v}\simeq 1$), while outside the void area we can observe some signal fluctuations 
that can be ascribed to the gas in the surrounding large-scale structures.
We remind that for each void we trim a local square $y$-map with the side equal to four times the void angular radius, in such a way as to contain the circular region of $2\times \theta_{\rm v}$ centred on the void;
this is also the outer edge to which we reconstruct the void profile. There is no standard recipe for the optimal outer radius in this kind of study; clearly it is possible to choose an arbitrarily large radius (e.g. $\geq 3\times \theta_{\rm v}$) to explore the signal behaviour outside the void radius. This, however, would come at the price of lowering the statistics of our sample, as only the voids that are fully overlapping with the ACT footprint over their full local (trimmed) square map are considered in the stack: extending the outer edge from $2\times \theta_{\rm v}$ to  $3\times \theta_{\rm v}$ will imply less voids can make this selection, leading to a lower S/N ratio in the final profile reconstruction. Besides, in this analysis we are mostly interested in the properties of the gas in the void inner region, and in principles, the void radius should mark the boundary of each void, as the former is associated with the corresponding circumspheres defined in the DIVE algorithm~\citep{Zhao2016}. For these reasons, in our analysis we choose $4\theta_{\rm v}$ as the side of the square stack map, i.e. $2\times \theta_{\rm v}$ as the radius of the circular region (we have also verified that beyond $2\theta_{\rm v}$ the signal amplitude drops back down to zero).

\subsection{Statistical characterisation} 

We estimate the covariance matrix $C_{ij}$ for the reconstructed $y$ profile employing the 1000 measured 
$y$ profiles obtained with the mock catalogues:
\begin{eqnarray}
	C_{ij} = \frac{1}{N_{\rm m}}\sum_{k=1}^{N_{\rm m}}\left(y_{i}^{(k)}- \langle y_{i}\rangle\right)\left(y_{j}^{(k)}-\langle y_{j}\rangle\right)
\end{eqnarray}
where $i,j$ index individual angular bins, $y_{i}^{(k)}$ is the measurement from the $k$-th mock, and $\langle y_{i} \rangle$ is the average across mocks. The correlation matrix, defined as
\begin{eqnarray}
R_{ij}=\frac{C_{ij}}{\sqrt{C_{ii}C_{jj}}}, \label{eq:def_corr}   
\end{eqnarray}
quantifies the level of correlation for off-diagonal bin pairs, and is plotted in 
Fig.~\ref{fig: Corre-Matrix}. We notice that the correlation matrices contain significant 
non-zero off-diagonal elements, which is likely a result of the beam smoothing in the Compton 
$y$-maps and also the mixing of scales occurring when rescaling the size of individual void maps 
before stacking. The uncertainties on the measured $y$ profiles can be quantified by the diagonal 
elements of the covariance matrix, yielding comparable error bars between ACT and \textit{Planck} cases.

To estimate the significance of our measurements, we compute the $\chi^2$ with a null hypothesis. 
In general, the total $\chi^2$ can be calculated as:
\begin{eqnarray}
    \chi^{2} = \sum_{i,j}\left(y_{i}-y_{i}^{\rm mod}\right)\,I_{ij}\,\left(y_{j}-y_{j}^{\rm mod}\right),
    \label{eq:chi2}
\end{eqnarray}
where $y_{i}$ is the measured Compton profile and $y_{i}^{\rm mod}$ is instead the given model; 
for the null hypothesis, $y^{\rm mod}=0$. The quantity $I_{ij}$ is the inverse covariance matrix, 
bias-corrected by included an extra factor as~\citep{Hartlap2007}:
\begin{eqnarray}
    I_{ij}=\frac{N_{\rm m}-N_{\rm b}-2}{N_{\rm m}-1}\left(C^{-1}\right)_{ij},
\end{eqnarray}
where $N_{\rm b}=20$ is the number of racial bins. 
We can also apply Eq.~\eqref{eq:chi2} to each mock profiles, which yields 1000 $\chi^2$ values 
for the null hypothesis; the distribution of these values, for both ACT and \textit{Planck}, is
plotted in Fig.~\ref{fig: chi2-mocks-v2}. This distribution should be described by a chi-squared 
distribution with 20 degrees of freedom, which is computed analytically and overplotted to 
Fig.~\ref{fig: chi2-mocks-v2}.
We find indeed a good match between 
the theoretical curve and the distributions obtained from the mock $\chi^2$ values. 

For the case of the real data stack, the $\chi^2$ for the null hypothesis ($y^{\rm mod}=0$) is evaluated at 
$\chi^{2}_{\rm ACT}=53.67$ for ACT and $\chi^{2}_{\it Planck}=93.91$ for \textit{Planck}, for the number of degree of freedom (dof) equal to $N_{\rm b}=20$. 
These correspond to the detection significance of $7.3\sigma$ for ACT and $9.7\sigma$ for \textit{Planck} measurements against the null signal.

We notice that, as far as the detection significance is concerned, there is no advantage in 
using ACT $y$-map over using {\it Planck} $y$-map, despite ACT considerably higher resolution. This could be 
ascribed primarily to the fact that voids are extended objects with mean radial angular sizes of 
$\theta_{v}\approx 1.2\,{\rm deg}$, implying that their signature is well resolved already with the resolution 
available in the \textit{Planck} map. In addition, we notice that, in fact, the smoothing effect caused by 
the larger \textit{Planck} beam typically determines lower background fluctuations in the random stacks, 
as it is visible in Fig.~\ref{fig: stack-results}, which results in \textit{Planck} error bars being 
slightly smaller than the ones obtained for ACT, as one can see from Fig.~\ref{fig:profiles}; 
this effect contributes to the slight higher value of \textit{Planck} significance when compared to ACT results.

It is worth mentioning, however, that~\citet{Alonso2018} only adopted a subsample of voids selected based on their high-significance identification by the void-finder algorithm, with the purpose of increasing the signal-to-noise (S/N) ratio of their measurements. For this reason, it is not meaningful to infer the expected S/N ratio with our sample purely based on the sample size. Another (this time technical) difference from the analysis in~\citet{Alonso2018} is that, while they perform a void stack on spherical coordinates ({\it Planck} $y$-map), we will perform it on a 2-D plane $y$-map (ACT). For a consistency check, we have verified that the stack of the same void sample used in~\citet{Alonso2018}, this time on the plane-projected \textit{Planck} map and using our stacking pipeline, provide results consistent with the ones presented by the authors.

\subsection{CIB contamination} 
\label{sect:systematic}

One of the main sources of systematic uncertainties comes from the cosmic infrared background 
(CIB;~\citealt{Planck-CIB2016}), that can contaminate the reconstructed tSZ $y$-maps. Following 
previous studies~\citep{Vikram2017,Hill2014,Alonso2018}, we use {\it Planck} 545 GHz temperature 
map (in units MJy sr$^{-1}$) as the proxy for CIB contamination, to quantify its potential influence on our 
tSZ measurement. The measured Compton signal from our stacks, $y_{\rm obs}$ can thus be modelled as 
a combination of the true, uncontaminated signal $y_{\rm true}$, and the CIB contribution, as:
\begin{eqnarray}
    y_{\rm obs}=y_{\rm true}+\alpha_{\rm CIB}T_{545},
\label{eq:cib}
\end{eqnarray}
where $T_{545}$ is the 545 GHz intensity map tracing the CIB, and $\alpha_{\rm CIB}$ is a scalar 
parameter gauging the CIB contribution. Here we take the fiducial value 
$\alpha_{\rm CIB}=(2.3\pm 6.6)\times 10^{-7}\,{\rm MJy}^{-1}\,{\rm sr}$, which was derived 
from the auto-correlation of the 545 GHz map and its cross-correlation with the {\it Planck} $y$ map in~\citet{Alonso2018}.  

In order to quantify the CIB contribution to our measured profiles, we can adopt the strategy in~\citet{Alonso2018} and stack the same set of voids on the $T_{545}$ map, to extract a 
mean $T_{545}$ void profile. The result is plotted in Fig.~\ref{fig: T545-profile}, where the lower 
amplitude at small separations can be interpreted as the cross-correlation signal between voids 
and the $T_{545}$ map. This profile is then scaled by $\alpha_{\rm CIB}$ following Eq.~\eqref{eq:cib}.

The signal leakage due to CIB (that is $\alpha_{\rm CIB}T_{545}$) into the reconstructed $y$ profiles
is shown with black diamonds in Fig.~\ref{fig: stack-results}, where we can see it is compatible with 
zero. The gray shaded region in Fig.~\ref{fig: stack-results} represents the 1-$\sigma$ uncertainty 
level of this leakage, which is computed by finding the maximum and minimum values of $\alpha_{\rm CIB}\times T_{545}(\theta)$ profile. One can see that this $\pm 1\sigma$ variation band is also much smaller than the measured tSZ signal, so the CIB residual in our void stacks is practically negligible. 
We further verified this conclusion by employing the 545 GHz CIB map from~\citep{Planck-CIB2016}, which 
separates the Galactic thermal dust emission from CIB 
anisotropies. The stack signal on this CIB map is also very small compared to the tSZ signal. Therefore, we conclude that the CIB contamination can be safely ignored in our stacking.

Finally, we comment on the effect of the telescope beam, which is convolved with the sky signal and 
may determine an amplitude dilution. However, we remind that the mean angular size of our voids is 
$\sim1.21$ deg, which is considerably more extended than the beam FWHM, even in the case of \textit{Planck}. 
Hence, possible systematic effects introduced by the instrumental beam are negligible in our analysis.


\section{Inference on gas properties}
\label{sect:gas}

In this section, we detail the theoretical model we adopt to describe our measured Compton 
void signal; specifically, we shall employ an isothermal model which assumes a uniform density of gas inside the void. 
The electron density $ n_{\rm e}({\bf x}, z)$ at any position $\mathbf{x}$ in space at 
redshift $z$ is related to the mean electron density $\Bar{n}_{\rm e}$ at the same redshift via:
\begin{eqnarray}
    n_{\rm e}({\bf x}, z)=\Bar{n}_{\rm e}(z)(1+\delta(\textbf{x})), \label{eq:ne}
\end{eqnarray}
where $\delta(\textbf{x})$ is the density contrast, and 
\begin{eqnarray}
    \Bar{n}_{\rm e}(z)=\frac{\rho_{\rm b}(z)}{\mu_{\rm e}m_{\rm p}} \label{eq:mean_ne}
\end{eqnarray}
is the cosmic mean electron density at redshift $z$~\citep{Ma2014}. In this relation 
$\mu_{\rm e}\simeq 1.14$ is the mean mass per proton, $m_{\rm p}$ is the proton mass, 
and $\rho_{\rm b}(z)=\rho_{\rm cr}\Omega_{\rm b}(1+z)^{3}$ is the baryon density at redshift $z$, 
where $\rho_{\rm cr}=1.879h^{2}\times 10^{-29}\,{\rm g}\,{\rm cm}^{-3}$ is the critical density of 
the Universe at present time, and $\Omega_{\rm b}$ is the baryon density parameter. In our modeling, we make the further simplifying assumption 
that the density contrast within the void is a constant, i.e. $\delta(\mathbf{x})\equiv \delta_{\rm v}$, which is expected to be negative.

We now calculate the measured Compton-$y$ profile of a void with a uniform electron temperature 
and density profile given by Eq.~(\ref{eq:ne}). Because in the stacking process we subtract the 
background term computed from random mock catalogues (Eq.~(\ref{eq:background_subtract})),  
the mean background $y$-value is effectively removed in the final observed $y(\theta)$ profiles 
plotted in Fig.~\ref{fig:profiles}.
We then subtract the same term in our theoretical modelling, effectively computing 
the contrast in the $y$ signal between voids and the mean background, as:
\begin{eqnarray}
y(r_{\perp}) &=& \frac{\sigma_{\rm T}k_{\rm B}}{m_{\rm e}c^{2}}\int \left[n_{\rm e}(\mathbf{x},z)-\bar{n}_{\rm e}(z) \right]T_{\rm e} \,{\rm d}l \nonumber \\
&=& \frac{\sigma_{\rm T}k_{\rm B}\bar{n}_{\rm e}(z)T_{\rm e}\delta_{\rm v}}{m_{\rm e}c^{2}}
    \int_{r_{\perp}}^{R_{\rm eff}}\frac{2r{\rm d}r}
    {\sqrt{r^{2}-r_{\perp}^{2}}} \nonumber \\
&=& \frac{2\sigma_{\rm T}k_{\rm B}\bar{n}_{\rm e}(z)T_{\rm e}\delta_{\rm v}}{m_{\rm e}c^{2}} \sqrt{R^{2}_{\rm eff}-r^{2}_{\perp}},  \label{eq:yperp} 
\end{eqnarray}
where $r_{\perp}$ is the projected distance from the void centre to the LoS, and $R_{\rm eff}$ is the void 
effective (physical) radius. In Eq.~(\ref{eq:yperp}), in the second equality, we changed the integration 
variable from the LoS distance $l$ to the void radial separation $r$, the two being related 
by $l\equiv \sqrt{r^{2}-r^{2}_{\perp}}$. We now substitute Eqs.~(\ref{eq:ne}) and (\ref{eq:mean_ne}) into Eq.~(\ref{eq:yperp}) and write the expression as a function of the angular separation $\theta$ from 
the void centre ($\theta=r_{\perp}/D_{\rm A}(z)$, $D_{\rm A}(z)$ being the angular diameter distance to the void effective redshift), which yields:
\begin{eqnarray}
y(\theta/\theta_{\rm v}) &=& \frac{2\sigma_{\rm T}k_{\rm B}T_{\rm e}R_{\rm v}\delta_{\rm v}}{m_{\rm e}c^{2}}\left(\frac{\Omega_{\rm b}\rho_{\rm cr}(1+z)^{3}}{\mu_{\rm e}m_{\rm p}} \right)\sqrt{1-\left(\frac{\theta}{\theta_{\rm v}} \right)^{2}}  \nonumber \\
&=& \left(7.1\times 10^{-10}\right)\delta_{\rm v}\left(\frac{T_{\rm e}}{10^{5}\,{\rm K}} \right)\sqrt{1-\left(\frac{\theta}{\theta_{\rm v}} \right)^{2}}, \label{eq:y-profile}
\end{eqnarray}
where $\theta_{\rm v}=R_{\rm eff}/D_{\rm A}(z)$, and in the last equality we have substituted 
the values for the void effective radius and redshift in our sample. 

\begin{figure*}
   \centering
   \includegraphics[width=8cm, angle=0]{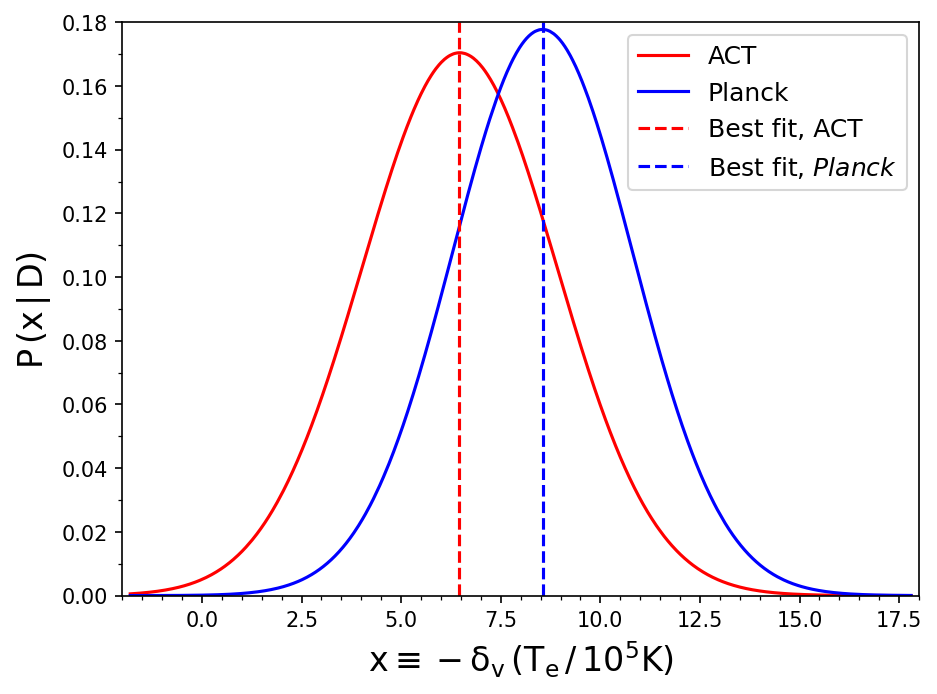}
   \includegraphics[width=8cm, angle=0]{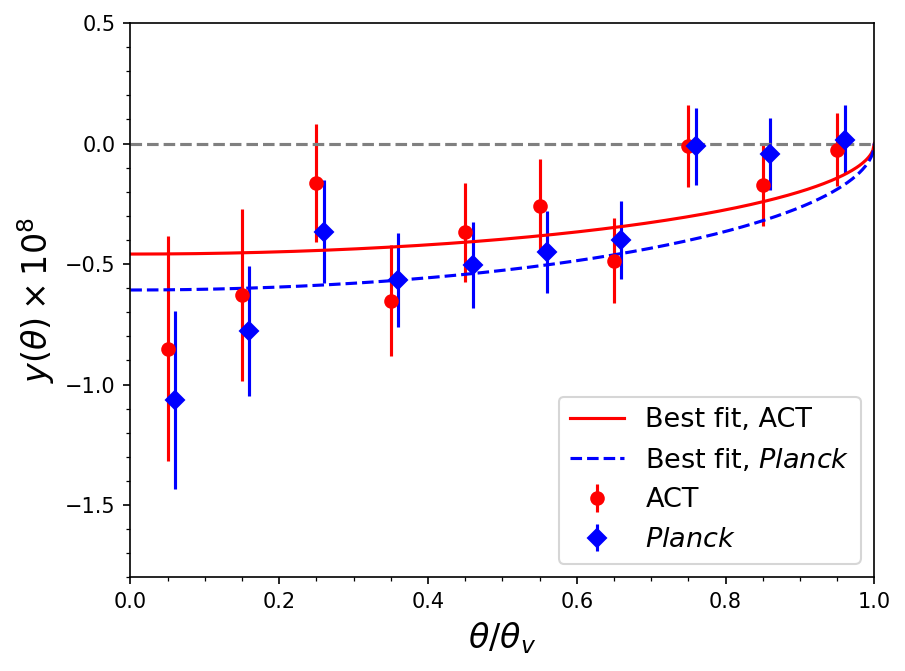}
   \caption{{\it Left}--the posterior probability distribution for the parameter 
	$(-\delta_{\rm v})(T_{\rm e}/10^{5}\,{\rm K})$ obtained from both ACT (solid red line)
	and {\it Planck} (dashed blue line) void stacking data; the vertical lines mark the position
	of their corresponding best-fit values. {\it Right}--comparison between 
	the best fit Compton-$y$ profiles (solid red line for ACT, dashed blue line for {\it Planck}) 
	and the stacked void data, as described in Sec.~\ref{sect:gas}.}
   \label{fig:bestfit}
\end{figure*}

The analytical $y$ profile predicted by Eq.~\eqref{eq:y-profile}, as a function of the scaled 
angular separation from the void centre $\theta/\theta_{\rm v}$, can be directly compared 
to our measured stacked profiles. As we do not know \textit{a priori} the values for the 
electron temperature and the 
density contrast inside the voids, we can leave the product $T_{\rm e}\delta_{\rm v}$ as a free 
parameter, and estimate its best-fit value against our data\footnote{As is shown in Eq.~\eqref{eq:y-profile}, the $T_{\rm e}$ and $\delta_{\rm v}$ in this model completely degenerate with each other, so it would not be meaningful to fit for the values of each of them separately.}. 
We sample the $(-\delta_{\rm v})(T_{\rm e}/10^{5}\,{\rm K})$ over the range of $[-2,18]$ for $1,000$ points, and calculate the likelihood from the $\chi^2$ in Eq.~\eqref{eq:chi2}; since we are only fitting for the void decrement signal, in order 
to not be biased by the neighboring structures we only limit the comparison between the data and the 
model prediction to the first $10$ data points (up to the void radius). The resulting posterior 
distribution on the parameter, for both the fits on ACT and \textit{Planck} data, is plotted in the 
left panel of Fig.~\ref{fig:bestfit}, whereas the right panel shows the comparison between the 
corresponding predictions and the measurements for the $y$ profiles. From the posterior distribution
we can extract the following best-fit estimates for the parameter $\delta_{\rm v}T_{\rm e}$:
\begin{eqnarray}
\delta T \equiv \left(-\delta_{\rm v}\frac{T_{\rm e}}{10^{5}\,{\rm K}} \right)=
\left\{\begin{array}{cc}
6.5 \pm 2.3\,(1\sigma) \pm 3.8\,(2\sigma) & {\rm ACT} \\
	8.6 \pm 2.1\,(1\sigma) \pm 3.5\,(2\sigma) & {\it Planck},
\end{array}
\right. 
\label{eq:deltaT}
\end{eqnarray}
where the negative sign accounts for the fact that we are fitting a decrement in the $y$ amplitude 
inside the void with respect to the mean background value. As for the goodness of this fit, it can 
be quantified by the $\chi^2$ computed using Eq.~\eqref{eq:chi2}, where this time $y^{\rm mod}$ is 
the predicted profile computed using Eq.~\eqref{eq:y-profile} and each of the best-fit values from 
Eq.~\eqref{eq:deltaT}. It is actually more meaningful to quote the reduced chi-squared $\chi^2_{\rm r}$, 
computed as the ratio between the chi-squared from Eq.~\eqref{eq:chi2} and the number of dof; the latter 
is equal to the number of points employed in the fit minus the number of fitted parameters, i.e. ${\rm dof}=10-1=9$. We then obtain $\chi^{2}_{\rm r, ACT}=1.71$  and $\chi^{2}_{\rm r,\it Planck}=2.47$ for ACT and \textit{Planck}, respectively. 
We notice that the reduced chi-square values suggest a non-optimal goodness of these fits. 
This can be expected in light of the simplified nature of our theoretical model, which may not capture equally 
well the gas properties at the void center and at the outer radii. We will come back to this issue in the 
next section.

Because voids are underdense regions, $\delta_{v}$ must satisfy $-1 \leq \delta_{v}< 0$; we can 
then infer lower limits on the gas temperature 
inside voids by saturating $\delta_{\rm v} \rightarrow -1$. By taking into account the uncertainties on our estimates, we find that the 
mean void electron temperature should satisfy $T_{\rm e}>2.7\times 10^{5}\,\text{K}$ for ACT and 
 $T_{\rm e}> 5.1\times 10^{5}\,\text{K}$ for \textit{Planck} (at 95\% C.L.).
 Results from hydrodynamical simulations~\citep{Martizzi2019} suggest that IGM and WHIM are the 
 major gas phases accounting for the total baryon budget, 
 spanning a broad temperature range from $10^{3}\,\text{K}$ to above 
 $10^{6}\,\text{K}$, which accommodates our findings.

We can also use our findings to provide estimates on the ratio between the electron number density 
in voids $n_{\rm e}^{\rm v}$ and its cosmic mean $\bar{n}_{\rm e}$, assuming uniform density of gas inside voids in Eq~\eqref{eq:ne}, we have:

\begin{eqnarray}
\frac{n_{\rm e}^{\rm v}}{\bar{n}_{\rm e}} &=& 1+\delta_{\rm v} \nonumber \\
&=& 1-\frac{\delta T \times 10^{5}{\rm K}}{T_{\rm e}} \nonumber \\
& \leqslant & 1-\frac{1}{10}\left(-\delta_{\rm v}\frac{T_{\rm e}}{10^{5}\,{\rm K}} \right) \nonumber \\
& \leqslant & 
\left\{\begin{array}{cc}
0.73 & 95\%\,\,{\rm C.L.}\,\, {\rm for}\,\,{\rm ACT}, \\
0.49 & \,\,95\%\,\,{\rm C.L.}\,\, {\rm for}\,\,{\it Planck},
\end{array}
\right. 
\label{eq:ne_v}
\end{eqnarray}
where the third inequality is taken by assuming the maximum gas temperature in 
voids to be $10^{6}\,\text{K}$ (the denominator in the second equality), as suggested by hydrodynamical simulations~\citep{Haider2016,Martizzi2019}. 
The fourth inequality is taken by substituting the $2\sigma$ lower limits from Eq.~(\ref{eq:deltaT}). Therefore, our lower limits in the void electron temperature then translate into upper limits 
for the void electron density, and where we find again 
broad agreement between the two data sets. 

\section{Conclusions}
\label{sect:conclusion}

In this paper, we have conducted a stacking analysis of a void catalogue constructed from BOSS
galaxy data, over the 1.6' resolution maps of the tSZ Compton $y$ parameter released by the ACT
collaboration, and also the {\it Planck} MILCA $y$-map (10' resolution) on the same footprint of ACT. We detected the void $y$-profile at $7.3\sigma$ for ACT and $9.7\sigma$ for {\it Planck}. Because the void signal we are targeting typically extends over degree angular scales 
on the sky, the higher resolution of ACT is not a major factor when determining the significance of 
the detection. Indeed, we observe similar features in both ACT and \textit{Planck} 2D stack maps, 
and similar trends in their corresponding radial $y$ profiles, confirming the consistency of our 
measurements. The uncertainties on the reconstructed profiles were computed by repeating the stacks 
on a set of 1000 mock catalogues, which also allowed us to evaluate the full covariance matrices 
associated with the profiles. Finally, we evaluated possible CIB contaminations in our stacks 
and found that correcting for a CIB leakage provides negligible changes in the profile amplitudes, 
well below the size of their error bars. 

We then attempted at reproducing the stacked profiles analytically by using a simple flat-density 
isothermal model for our voids. Although these are strong assumptions, this is intended to be a 
preliminary exploration of how the profiles reconstructed from tSZ data can constrain the gas 
properties within voids. Our model leaves the product between the density contrast and temperature 
of gas within voids as a free paramter, which we fitted against our measurements to 
$-\delta_{\rm v}\times T_{\rm e}=\left(6.5 \pm 2.3 \right)\times 10^{5} \,\text{K}$ using ACT data and 
$-\delta_{\rm v}\times T_{\rm e}=\left(8.6 \pm 2.1 \right)\times 10^{5} \,\text{K}$ using \textit{Planck} 
data, with a remarkable agreement between the two data sets. 
These constraints allowed us to place lower limits on the void gas temperature, at 
$2.7 \times10^{5}\,\text{K}$ for ACT and $5.1 \times10^{5}\,\text{K}$ for \textit{Planck} 
($95\%$ C.L.), which fall relatively close to the upper limit of the range spanned by numerical 
simulation results, suggesting voids may be warmer than expected. An increase in the void gas temperature 
can be induced by feedback processes from galaxy hosting active galactic nuclei (AGNs), 
either located inside the void or in its neighbouring region and ejecting hot gas towards the void 
centre~\citep{Constantin2008,Haider2016,Martizzi2019}. 
Our lower limits on $T_{\rm e}$ enabled us to assess upper limits on the ratio between the 
electron number density within voids and its cosmic mean value; the results suggest a 
deficit of electrons in the void region. Indeed, by setting $10^{6}\,\text{K}$ as the highest 
temperature inside voids, an assumption derived from numerical simulations~\citep{Martizzi2019}, 
the computed fraction of electron number density against the cosmic mean is expected to be 
below $0.73$ (ACT) or $0.49$ (\textit{Planck}) at $95\%$ C.L. This results corroborates the finding 
that voids, as under-dense cosmic structures, are under-pressured relative to the cosmic 
mean~\citep{Alonso2018}. The electron number density in our model is a constant in the voids, which is unrelated with the gas temperature in the voids. We notice that in~\citet{Bolton2008},  an inverted temperature-density relation is proposed (the middle panel in their Fig.~3), for gas at redshift $z\simeq 2$ with temperature $T_{\rm e}<10^{5}\,{\rm K}$. But because our work and \citet{Bolton2008} are dealing with gas in different redshifts and temperature regimes, the gas and temperature relation is not always true across different physical conditions. We do notice that one can test this relation once the stacking significance is further improved.

However, we acknowledge that our theoretical model is quite simple and may not be very 
accurate in capturing the electron density and temperature variations within voids. In fact, the void 
density may experience significant transitions from its center to its 
edge~\citep{Hamaus2014,Chantavat2017,Verza2022}. Besides, the general linear relation between electron number density and the dark matter density 
profile may not be accurate in under-dense environments, possibly requiring a position-dependent bias. 
Finally, the temperature in voids can deviate from the isothermal profile with a complex radial 
dependence because of the presence of multi-phase gas, as revealed recently by 
simulations~\citep{Haider2016,Martizzi2019}. Still, because the main goal of our current work is an observational 
detection of the void signal, we did not explore more detailed models.  We 
leave the implementation of a more accurate theoretical model to a future dedicated study, once the significance of the measurement is further improved.

As a final remark, from an observational perspective, the tSZ signal around voids is generally 
weak and sensitive to the choice of the void catalogue and the tSZ map. In the future, an 
optimal reconstruction of high-resolution, large area tSZ signal, with a better cleaning of 
potential contamination sources, in combination with a larger number of well-selected voids, 
can undoubtedly improve the results of this type of analysis.

\section*{Acknowledgements}

We thank useful discussion with David Alonso, Cheng Zhao and Yi-Peng Jing. 
YZM acknowledges the support of National Research Foundation with Grant No. 150385, and the research program ``New Insights into Astrophysics and Cosmology with Theoretical Models Confronting Observational Data'' of the National Institute for Theoretical and Computational Sciences of South Africa. DT acknowledges financial support from the XJTLU Research Development Fund (RDF) grant with number RDF-22-02-068. We acknowledge the cosmology simulation database (CSD) in the National Basic Science 
Data Center (NBSDC) with the funding NBSDC-DB-10.

\section*{Data Availability}
The {\it Planck} $y$-map can be downloaded from the {\it Planck} Legacy Archive 
website\footnote{\url{http://pla.esac.esa.int/pla/}}, whereas the ACT $y$-map can be 
downloaded from the NASA LAMBDA website\footnote{\url{https://lambda.gsfc.nasa.gov/product/act/}}. 
The other data products employed in this work can be delivered for verification purposes 
under request to the first and corresponding authors.



\bibliographystyle{mnras}
\bibliography{example} 

\begin{thebibliography}{}
\makeatletter
\relax
\def\mn@urlcharsother{\let\do\@makeother \do\$\do\&\do\#\do\^\do\_\do\%\do\~}
\def\mn@doi{\begingroup\mn@urlcharsother \@ifnextchar [ {\mn@doi@}
  {\mn@doi@[]}}
\def\mn@doi@[#1]#2{\def\@tempa{#1}\ifx\@tempa\@empty \href
  {http://dx.doi.org/#2} {doi:#2}\else \href {http://dx.doi.org/#2} {#1}\fi
  \endgroup}
\def\mn@eprint#1#2{\mn@eprint@#1:#2::\@nil}
\def\mn@eprint@arXiv#1{\href {http://arxiv.org/abs/#1} {{\tt arXiv:#1}}}
\def\mn@eprint@dblp#1{\href {http://dblp.uni-trier.de/rec/bibtex/#1.xml}
  {dblp:#1}}
\def\mn@eprint@#1:#2:#3:#4\@nil{\def\@tempa {#1}\def\@tempb {#2}\def\@tempc
  {#3}\ifx \@tempc \@empty \let \@tempc \@tempb \let \@tempb \@tempa \fi \ifx
  \@tempb \@empty \def\@tempb {arXiv}\fi \@ifundefined
  {mn@eprint@\@tempb}{\@tempb:\@tempc}{\expandafter \expandafter \csname
  mn@eprint@\@tempb\endcsname \expandafter{\@tempc}}}

\bibitem[\protect\citeauthoryear{{Alam} et~al.,}{{Alam}
  et~al.}{2017}]{Alam2017}
{Alam} S.,  et~al., 2017, \mn@doi [\mnras] {10.1093/mnras/stx721}, \href
  {https://ui.adsabs.harvard.edu/abs/2017MNRAS.470.2617A} {470, 2617}

\bibitem[\protect\citeauthoryear{{Alonso}, {Hill}, {Hlo{\v{z}}ek}  \&
  {Spergel}}{{Alonso} et~al.}{2018}]{Alonso2018}
{Alonso} D.,  {Hill} J.~C.,  {Hlo{\v{z}}ek} R.,   {Spergel} D.~N.,  2018,
  \mn@doi [\prd] {10.1103/PhysRevD.97.063514}, \href
  {https://ui.adsabs.harvard.edu/abs/2018PhRvD..97f3514A} {97, 063514}

\bibitem[\protect\citeauthoryear{{Aubert} et~al.,}{{Aubert}
  et~al.}{2022}]{Aubert2022}
{Aubert} M.,  et~al., 2022, \mn@doi [\mnras] {10.1093/mnras/stac828}, \href
  {https://ui.adsabs.harvard.edu/abs/2022MNRAS.513..186A} {513, 186}

\bibitem[\protect\citeauthoryear{{Beygu}, {Peletier}, {van der Hulst},
  {Jarrett}, {Kreckel}, {van de Weygaert}, {van Gorkom}  \&
  {Aragon-Calvo}}{{Beygu} et~al.}{2017}]{Beygu2017}
{Beygu} B.,  {Peletier} R.~F.,  {van der Hulst} J.~M.,  {Jarrett} T.~H.,
  {Kreckel} K.,  {van de Weygaert} R.,  {van Gorkom} J.~H.,   {Aragon-Calvo}
  M.~A.,  2017, \mn@doi [\mnras] {10.1093/mnras/stw2362}, \href
  {https://ui.adsabs.harvard.edu/abs/2017MNRAS.464..666B} {464, 666}

\bibitem[\protect\citeauthoryear{{Bolton}, {Viel}, {Kim}, {Haehnelt}  \&
  {Carswell}}{{Bolton} et~al.}{2008}]{Bolton2008}
{Bolton} J.~S.,  {Viel} M.,  {Kim} T.~S.,  {Haehnelt} M.~G.,   {Carswell}
  R.~F.,  2008, \mn@doi [\mnras] {10.1111/j.1365-2966.2008.13114.x}, \href
  {https://ui.adsabs.harvard.edu/abs/2008MNRAS.386.1131B} {386, 1131}

\bibitem[\protect\citeauthoryear{{Bonjean}, {Aghanim}, {Salom{\'e}}, {Douspis}
  \& {Beelen}}{{Bonjean} et~al.}{2018}]{Bonjean2018}
{Bonjean} V.,  {Aghanim} N.,  {Salom{\'e}} P.,  {Douspis} M.,   {Beelen} A.,
  2018, \mn@doi [\aap] {10.1051/0004-6361/201731699}, \href
  {https://ui.adsabs.harvard.edu/abs/2018A&A...609A..49B} {609, A49}

\bibitem[\protect\citeauthoryear{{Bregman}}{{Bregman}}{2007}]{Bregman2007}
{Bregman} J.~N.,  2007, \mn@doi [\araa]
  {10.1146/annurev.astro.45.051806.110619}, \href
  {https://ui.adsabs.harvard.edu/abs/2007ARA&A..45..221B} {45, 221}

\bibitem[\protect\citeauthoryear{{Cai}, {Padilla}  \& {Li}}{{Cai}
  et~al.}{2015}]{Cai2015}
{Cai} Y.-C.,  {Padilla} N.,   {Li} B.,  2015, \mn@doi [\mnras]
  {10.1093/mnras/stv777}, \href
  {https://ui.adsabs.harvard.edu/abs/2015MNRAS.451.1036C} {451, 1036}

\bibitem[\protect\citeauthoryear{{Cai}, {Neyrinck}, {Mao}, {Peacock}, {Szapudi}
   \& {Berlind}}{{Cai} et~al.}{2017}]{Cai2017}
{Cai} Y.-C.,  {Neyrinck} M.,  {Mao} Q.,  {Peacock} J.~A.,  {Szapudi} I.,
  {Berlind} A.~A.,  2017, \mn@doi [\mnras] {10.1093/mnras/stw3299}, \href
  {https://ui.adsabs.harvard.edu/abs/2017MNRAS.466.3364C} {466, 3364}

\bibitem[\protect\citeauthoryear{{Ceccarelli}, {Duplancic}  \& {Garcia
  Lambas}}{{Ceccarelli} et~al.}{2022}]{Ceccarelli2022}
{Ceccarelli} L.,  {Duplancic} F.,   {Garcia Lambas} D.,  2022, \mn@doi [\mnras]
  {10.1093/mnras/stab2902}, \href
  {https://ui.adsabs.harvard.edu/abs/2022MNRAS.509.1805C} {509, 1805}

\bibitem[\protect\citeauthoryear{{Chantavat}, {Sawangwit}  \&
  {Wandelt}}{{Chantavat} et~al.}{2017}]{Chantavat2017}
{Chantavat} T.,  {Sawangwit} U.,   {Wandelt} B.~D.,  2017, \mn@doi [\apj]
  {10.3847/1538-4357/836/2/156}, \href
  {https://ui.adsabs.harvard.edu/abs/2017ApJ...836..156C} {836, 156}

\bibitem[\protect\citeauthoryear{{Clampitt} \& {Jain}}{{Clampitt} \&
  {Jain}}{2015}]{Clampitt2015}
{Clampitt} J.,  {Jain} B.,  2015, \mn@doi [\mnras] {10.1093/mnras/stv2215},
  \href {https://ui.adsabs.harvard.edu/abs/2015MNRAS.454.3357C} {454, 3357}

\bibitem[\protect\citeauthoryear{{Constantin}, {Hoyle}  \&
  {Vogeley}}{{Constantin} et~al.}{2008}]{Constantin2008}
{Constantin} A.,  {Hoyle} F.,   {Vogeley} M.~S.,  2008, \mn@doi [\apj]
  {10.1086/524310}, \href
  {https://ui.adsabs.harvard.edu/abs/2008ApJ...673..715C} {673, 715}

\bibitem[\protect\citeauthoryear{{Falck}, {Koyama}, {Zhao}  \&
  {Cautun}}{{Falck} et~al.}{2018}]{Falck2018}
{Falck} B.,  {Koyama} K.,  {Zhao} G.-B.,   {Cautun} M.,  2018, \mn@doi [\mnras]
  {10.1093/mnras/stx3288}, \href
  {https://ui.adsabs.harvard.edu/abs/2018MNRAS.475.3262F} {475, 3262}

\bibitem[\protect\citeauthoryear{{Fang} et~al.,}{{Fang}
  et~al.}{2019}]{Fang2019}
{Fang} Y.,  et~al., 2019, \mn@doi [\mnras] {10.1093/mnras/stz2805}, \href
  {https://ui.adsabs.harvard.edu/abs/2019MNRAS.490.3573F} {490, 3573}

\bibitem[\protect\citeauthoryear{{Fukugita} \& {Peebles}}{{Fukugita} \&
  {Peebles}}{2004}]{Fukugita2004}
{Fukugita} M.,  {Peebles} P.~J.~E.,  2004, \mn@doi [\apj] {10.1086/425155},
  \href {https://ui.adsabs.harvard.edu/abs/2004ApJ...616..643F} {616, 643}

\bibitem[\protect\citeauthoryear{{G{\'o}rski}, {Hivon}, {Banday}, {Wandelt},
  {Hansen}, {Reinecke}  \& {Bartelmann}}{{G{\'o}rski}
  et~al.}{2005}]{Gorski2005}
{G{\'o}rski} K.~M.,  {Hivon} E.,  {Banday} A.~J.,  {Wandelt} B.~D.,  {Hansen}
  F.~K.,  {Reinecke} M.,   {Bartelmann} M.,  2005, \mn@doi [\apj]
  {10.1086/427976}, \href
  {https://ui.adsabs.harvard.edu/abs/2005ApJ...622..759G} {622, 759}

\bibitem[\protect\citeauthoryear{{Haider}, {Steinhauser}, {Vogelsberger},
  {Genel}, {Springel}, {Torrey}  \& {Hernquist}}{{Haider}
  et~al.}{2016}]{Haider2016}
{Haider} M.,  {Steinhauser} D.,  {Vogelsberger} M.,  {Genel} S.,  {Springel}
  V.,  {Torrey} P.,   {Hernquist} L.,  2016, \mn@doi [\mnras]
  {10.1093/mnras/stw077}, \href
  {https://ui.adsabs.harvard.edu/abs/2016MNRAS.457.3024H} {457, 3024}

\bibitem[\protect\citeauthoryear{{Hamaus}, {Sutter}  \& {Wandelt}}{{Hamaus}
  et~al.}{2014}]{Hamaus2014}
{Hamaus} N.,  {Sutter} P.~M.,   {Wandelt} B.~D.,  2014, \mn@doi [\prl]
  {10.1103/PhysRevLett.112.251302}, \href
  {https://ui.adsabs.harvard.edu/abs/2014PhRvL.112y1302H} {112, 251302}

\bibitem[\protect\citeauthoryear{{Hamaus}, {Pisani}, {Choi}, {Lavaux},
  {Wandelt}  \& {Weller}}{{Hamaus} et~al.}{2020}]{Hamaus2020}
{Hamaus} N.,  {Pisani} A.,  {Choi} J.-A.,  {Lavaux} G.,  {Wandelt} B.~D.,
  {Weller} J.,  2020, \mn@doi [\jcap] {10.1088/1475-7516/2020/12/023}, \href
  {https://ui.adsabs.harvard.edu/abs/2020JCAP...12..023H} {2020, 023}

\bibitem[\protect\citeauthoryear{{Hartlap}, {Simon}  \& {Schneider}}{{Hartlap}
  et~al.}{2007}]{Hartlap2007}
{Hartlap} J.,  {Simon} P.,   {Schneider} P.,  2007, \mn@doi [\aap]
  {10.1051/0004-6361:20066170}, \href
  {https://ui.adsabs.harvard.edu/abs/2007A&A...464..399H} {464, 399}

\bibitem[\protect\citeauthoryear{{Hill} \& {Spergel}}{{Hill} \&
  {Spergel}}{2014}]{Hill2014}
{Hill} J.~C.,  {Spergel} D.~N.,  2014, \mn@doi [\jcap]
  {10.1088/1475-7516/2014/02/030}, \href
  {https://ui.adsabs.harvard.edu/abs/2014JCAP...02..030H} {2014, 030}

\bibitem[\protect\citeauthoryear{{Hojjati}, {McCarthy}, {Harnois-Deraps}, {Ma},
  {Waerbeke}, {Hinshaw}  \& {Brun}}{{Hojjati} et~al.}{2015}]{Hojjati2015}
{Hojjati} A.,  {McCarthy} I.~G.,  {Harnois-Deraps} J.,  {Ma} Y.-Z.,  {Waerbeke}
  L.~V.,  {Hinshaw} G.,   {Brun} A. M.~C.~L.,  2015, \mn@doi [\jcap]
  {10.1088/1475-7516/2015/10/047}, \href
  {https://ui.adsabs.harvard.edu/abs/2015JCAP...10..047H} {2015, 047}

\bibitem[\protect\citeauthoryear{{Hojjati} et~al.,}{{Hojjati}
  et~al.}{2017}]{Hojjati2017}
{Hojjati} A.,  et~al., 2017, \mn@doi [\mnras] {10.1093/mnras/stx1659}, \href
  {https://ui.adsabs.harvard.edu/abs/2017MNRAS.471.1565H} {471, 1565}

\bibitem[\protect\citeauthoryear{{Hurier}, {Mac{\'\i}as-P{\'e}rez}  \&
  {Hildebrandt}}{{Hurier} et~al.}{2013}]{Hurier2013}
{Hurier} G.,  {Mac{\'\i}as-P{\'e}rez} J.~F.,   {Hildebrandt} S.,  2013, \mn@doi
  [\aap] {10.1051/0004-6361/201321891}, \href
  {https://ui.adsabs.harvard.edu/abs/2013A&A...558A.118H} {558, A118}

\bibitem[\protect\citeauthoryear{{Ibitoye}, {Tramonte}, {Ma}  \&
  {Dai}}{{Ibitoye} et~al.}{2022}]{Ibitoye2022}
{Ibitoye} A.,  {Tramonte} D.,  {Ma} Y.-Z.,   {Dai} W.-M.,  2022, \mn@doi [\apj]
  {10.3847/1538-4357/ac7b8c}, \href
  {https://ui.adsabs.harvard.edu/abs/2022ApJ...935...18I} {935, 18}

\bibitem[\protect\citeauthoryear{{Lavaux} \& {Wandelt}}{{Lavaux} \&
  {Wandelt}}{2012}]{Lavaux2012}
{Lavaux} G.,  {Wandelt} B.~D.,  2012, \mn@doi [\apj]
  {10.1088/0004-637X/754/2/109}, \href
  {https://ui.adsabs.harvard.edu/abs/2012ApJ...754..109L} {754, 109}

\bibitem[\protect\citeauthoryear{{Ma} \& {Zhao}}{{Ma} \& {Zhao}}{2014}]{Ma2014}
{Ma} Y.-Z.,  {Zhao} G.-B.,  2014, \mn@doi [Physics Letters B]
  {10.1016/j.physletb.2014.06.066}, \href
  {https://ui.adsabs.harvard.edu/abs/2014PhLB..735..402M} {735, 402}

\bibitem[\protect\citeauthoryear{{Ma}, {Waerbeke}, {Hinshaw}, {Hojjati},
  {Scott}  \& {Zuntz}}{{Ma} et~al.}{2015}]{Ma2015}
{Ma} Y.-Z.,  {Waerbeke} L.~V.,  {Hinshaw} G.,  {Hojjati} A.,  {Scott} D.,
  {Zuntz} J.,  2015, \mn@doi [\jcap] {10.1088/1475-7516/2015/09/046}, \href
  {https://ui.adsabs.harvard.edu/abs/2015JCAP...09..046M} {2015, 046}

\bibitem[\protect\citeauthoryear{{Ma}, {Gong}, {Tr{\"o}ster}  \& {Van
  Waerbeke}}{{Ma} et~al.}{2021}]{Ma2021}
{Ma} Y.-Z.,  {Gong} Y.,  {Tr{\"o}ster} T.,   {Van Waerbeke} L.,  2021, \mn@doi
  [\mnras] {10.1093/mnras/staa3369}, \href
  {https://ui.adsabs.harvard.edu/abs/2021MNRAS.500.1806M} {500, 1806}

\bibitem[\protect\citeauthoryear{{Macquart} et~al.,}{{Macquart}
  et~al.}{2020}]{Macquart2020}
{Macquart} J.~P.,  et~al., 2020, \mn@doi [\nat] {10.1038/s41586-020-2300-2},
  \href {https://ui.adsabs.harvard.edu/abs/2020Natur.581..391M} {581, 391}

\bibitem[\protect\citeauthoryear{{Madhavacheril} et~al.,}{{Madhavacheril}
  et~al.}{2020}]{Madhavacheril2020}
{Madhavacheril} M.~S.,  et~al., 2020, \mn@doi [\prd]
  {10.1103/PhysRevD.102.023534}, \href
  {https://ui.adsabs.harvard.edu/abs/2020PhRvD.102b3534M} {102, 023534}

\bibitem[\protect\citeauthoryear{{Makiya}, {Ando}  \& {Komatsu}}{{Makiya}
  et~al.}{2018}]{Makiya2018}
{Makiya} R.,  {Ando} S.,   {Komatsu} E.,  2018, \mn@doi [\mnras]
  {10.1093/mnras/sty2031}, \href
  {https://ui.adsabs.harvard.edu/abs/2018MNRAS.480.3928M} {480, 3928}

\bibitem[\protect\citeauthoryear{{Martizzi} et~al.,}{{Martizzi}
  et~al.}{2019}]{Martizzi2019}
{Martizzi} D.,  et~al., 2019, \mn@doi [\mnras] {10.1093/mnras/stz1106}, \href
  {https://ui.adsabs.harvard.edu/abs/2019MNRAS.486.3766M} {486, 3766}

\bibitem[\protect\citeauthoryear{{Melchior}, {Sutter}, {Sheldon}, {Krause}  \&
  {Wandelt}}{{Melchior} et~al.}{2014}]{Melchior2014}
{Melchior} P.,  {Sutter} P.~M.,  {Sheldon} E.~S.,  {Krause} E.,   {Wandelt}
  B.~D.,  2014, \mn@doi [\mnras] {10.1093/mnras/stu456}, \href
  {https://ui.adsabs.harvard.edu/abs/2014MNRAS.440.2922M} {440, 2922}

\bibitem[\protect\citeauthoryear{{Nicastro} et~al.,}{{Nicastro}
  et~al.}{2018}]{Nicastro2018}
{Nicastro} F.,  et~al., 2018, \mn@doi [\nat] {10.1038/s41586-018-0204-1}, \href
  {https://ui.adsabs.harvard.edu/abs/2018Natur.558..406N} {558, 406}

\bibitem[\protect\citeauthoryear{{Pandey} et~al.,}{{Pandey}
  et~al.}{2022}]{Pandey2022}
{Pandey} S.,  et~al., 2022, \mn@doi [\prd] {10.1103/PhysRevD.105.123526}, \href
  {https://ui.adsabs.harvard.edu/abs/2022PhRvD.105l3526P} {105, 123526}

\bibitem[\protect\citeauthoryear{{Pisani}, {Sutter}, {Hamaus}, {Alizadeh},
  {Biswas}, {Wandelt}  \& {Hirata}}{{Pisani} et~al.}{2015}]{Pisani2015}
{Pisani} A.,  {Sutter} P.~M.,  {Hamaus} N.,  {Alizadeh} E.,  {Biswas} R.,
  {Wandelt} B.~D.,   {Hirata} C.~M.,  2015, \mn@doi [\prd]
  {10.1103/PhysRevD.92.083531}, \href
  {https://ui.adsabs.harvard.edu/abs/2015PhRvD..92h3531P} {92, 083531}

\bibitem[\protect\citeauthoryear{{Planck Collaboration} et~al.,}{{Planck
  Collaboration} et~al.}{2013}]{Planck2013}
{Planck Collaboration} et~al., 2013, \mn@doi [\aap]
  {10.1051/0004-6361/201220194}, \href
  {https://ui.adsabs.harvard.edu/abs/2013A&A...550A.134P} {550, A134}

\bibitem[\protect\citeauthoryear{{Planck Collaboration} et~al.,}{{Planck
  Collaboration} et~al.}{2016a}]{Planck-ymap2016}
{Planck Collaboration} et~al., 2016a, \mn@doi [\aap]
  {10.1051/0004-6361/201525826}, \href
  {https://ui.adsabs.harvard.edu/abs/2016A&A...594A..22P} {594, A22}

\bibitem[\protect\citeauthoryear{{Planck Collaboration} et~al.,}{{Planck
  Collaboration} et~al.}{2016b}]{Planck-CIB2016}
{Planck Collaboration} et~al., 2016b, \mn@doi [\aap]
  {10.1051/0004-6361/201527418}, \href
  {https://ui.adsabs.harvard.edu/abs/2016A&A...594A..23P} {594, A23}

\bibitem[\protect\citeauthoryear{{Planck Collaboration} et~al.,}{{Planck
  Collaboration} et~al.}{2016c}]{Planck2016SZ-sources}
{Planck Collaboration} et~al., 2016c, \mn@doi [\aap]
  {10.1051/0004-6361/201525823}, \href
  {https://ui.adsabs.harvard.edu/abs/2016A&A...594A..27P} {594, A27}

\bibitem[\protect\citeauthoryear{{Planck Collaboration} et~al.,}{{Planck
  Collaboration} et~al.}{2020}]{Planck-2018parameters}
{Planck Collaboration} et~al., 2020, \mn@doi [\aap]
  {10.1051/0004-6361/201833910}, \href
  {https://ui.adsabs.harvard.edu/abs/2020A&A...641A...6P} {641, A6}

\bibitem[\protect\citeauthoryear{{Pustilnik}, {Tepliakova}  \&
  {Makarov}}{{Pustilnik} et~al.}{2019}]{Pustilnik2019}
{Pustilnik} S.~A.,  {Tepliakova} A.~L.,   {Makarov} D.~I.,  2019, \mn@doi
  [\mnras] {10.1093/mnras/sty2947}, \href
  {https://ui.adsabs.harvard.edu/abs/2019MNRAS.482.4329P} {482, 4329}

\bibitem[\protect\citeauthoryear{{Raghunathan}, {Nadathur}, {Sherwin}  \&
  {Whitehorn}}{{Raghunathan} et~al.}{2020}]{Raghunathan2020}
{Raghunathan} S.,  {Nadathur} S.,  {Sherwin} B.~D.,   {Whitehorn} N.,  2020,
  \mn@doi [\apj] {10.3847/1538-4357/ab6f05}, \href
  {https://ui.adsabs.harvard.edu/abs/2020ApJ...890..168R} {890, 168}

\bibitem[\protect\citeauthoryear{{Remazeilles}, {Delabrouille}  \&
  {Cardoso}}{{Remazeilles} et~al.}{2011}]{Remazeilles2011}
{Remazeilles} M.,  {Delabrouille} J.,   {Cardoso} J.-F.,  2011, \mn@doi
  [\mnras] {10.1111/j.1365-2966.2010.17624.x}, \href
  {https://ui.adsabs.harvard.edu/abs/2011MNRAS.410.2481R} {410, 2481}

\bibitem[\protect\citeauthoryear{{S{\'a}nchez} et~al.,}{{S{\'a}nchez}
  et~al.}{2017}]{Sanchez2017}
{S{\'a}nchez} C.,  et~al., 2017, \mn@doi [\mnras] {10.1093/mnras/stw2745},
  \href {https://ui.adsabs.harvard.edu/abs/2017MNRAS.465..746S} {465, 746}

\bibitem[\protect\citeauthoryear{{Shull}, {Smith}  \& {Danforth}}{{Shull}
  et~al.}{2012}]{Shull2012}
{Shull} J.~M.,  {Smith} B.~D.,   {Danforth} C.~W.,  2012, \mn@doi [\apj]
  {10.1088/0004-637X/759/1/23}, \href
  {https://ui.adsabs.harvard.edu/abs/2012ApJ...759...23S} {759, 23}

\bibitem[\protect\citeauthoryear{{Sunyaev} \& {Zeldovich}}{{Sunyaev} \&
  {Zeldovich}}{1970}]{Sunyaev1970}
{Sunyaev} R.~A.,  {Zeldovich} Y.~B.,  1970, \mn@doi [\apss]
  {10.1007/BF00653471}, \href
  {https://ui.adsabs.harvard.edu/abs/1970Ap&SS...7....3S} {7, 3}

\bibitem[\protect\citeauthoryear{{Tanidis} \& {Camera}}{{Tanidis} \&
  {Camera}}{2021}]{Tanidis2021}
{Tanidis} K.,  {Camera} S.,  2021, \mn@doi [arXiv e-prints]
  {10.48550/arXiv.2107.00026}, \href
  {https://ui.adsabs.harvard.edu/abs/2021arXiv210700026T} {p. arXiv:2107.00026}

\bibitem[\protect\citeauthoryear{{Tanimura}, {Aghanim}, {Douspis}, {Beelen}  \&
  {Bonjean}}{{Tanimura} et~al.}{2019}]{Tanimura2019}
{Tanimura} H.,  {Aghanim} N.,  {Douspis} M.,  {Beelen} A.,   {Bonjean} V.,
  2019, \mn@doi [\aap] {10.1051/0004-6361/201833413}, \href
  {https://ui.adsabs.harvard.edu/abs/2019A&A...625A..67T} {625, A67}

\bibitem[\protect\citeauthoryear{{Tramonte} et~al.,}{{Tramonte}
  et~al.}{2023}]{Tramonte2023}
{Tramonte} D.,  et~al., 2023, \mn@doi [\apjs] {10.3847/1538-4365/acbcca}, \href
  {https://ui.adsabs.harvard.edu/abs/2023ApJS..265...55T} {265, 55}

\bibitem[\protect\citeauthoryear{{Verza}, {Carbone}  \& {Renzi}}{{Verza}
  et~al.}{2022}]{Verza2022}
{Verza} G.,  {Carbone} C.,   {Renzi} A.,  2022, \mn@doi [\apjl]
  {10.3847/2041-8213/ac9d98}, \href
  {https://ui.adsabs.harvard.edu/abs/2022ApJ...940L..16V} {940, L16}

\bibitem[\protect\citeauthoryear{{Vielzeuf} et~al.,}{{Vielzeuf}
  et~al.}{2021}]{Vielzeuf2021}
{Vielzeuf} P.,  et~al., 2021, \mn@doi [\mnras] {10.1093/mnras/staa3231}, \href
  {https://ui.adsabs.harvard.edu/abs/2021MNRAS.500..464V} {500, 464}

\bibitem[\protect\citeauthoryear{{Vikram}, {Lidz}  \& {Jain}}{{Vikram}
  et~al.}{2017}]{Vikram2017}
{Vikram} V.,  {Lidz} A.,   {Jain} B.,  2017, \mn@doi [\mnras]
  {10.1093/mnras/stw3311}, \href
  {https://ui.adsabs.harvard.edu/abs/2017MNRAS.467.2315V} {467, 2315}

\bibitem[\protect\citeauthoryear{{Zhao}, {Tao}, {Liang}, {Kitaura}  \&
  {Chuang}}{{Zhao} et~al.}{2016}]{Zhao2016}
{Zhao} C.,  {Tao} C.,  {Liang} Y.,  {Kitaura} F.-S.,   {Chuang} C.-H.,  2016,
  \mn@doi [\mnras] {10.1093/mnras/stw660}, \href
  {https://ui.adsabs.harvard.edu/abs/2016MNRAS.459.2670Z} {459, 2670}

\bibitem[\protect\citeauthoryear{{de Graaff}, {Cai}, {Heymans}  \&
  {Peacock}}{{de Graaff} et~al.}{2019}]{deGraaff2019}
{de Graaff} A.,  {Cai} Y.-C.,  {Heymans} C.,   {Peacock} J.~A.,  2019, \mn@doi
  [\aap] {10.1051/0004-6361/201935159}, \href
  {https://ui.adsabs.harvard.edu/abs/2019A&A...624A..48D} {624, A48}

\makeatother
\end{thebibliography}

\bsp	
\label{lastpage}
\end{document}